\documentclass[journal,comsoc]{IEEEtran}
\IEEEoverridecommandlockouts

\usepackage{float} 
\usepackage{makecell}
\usepackage{multirow}
\usepackage{cite}
\usepackage{amsmath,amssymb,amsfonts}
\usepackage{algorithmic}
\usepackage{graphicx}
\usepackage{textcomp}
\usepackage{xcolor}
\usepackage{url}
\usepackage{booktabs}
\usepackage{hyperref}
\usepackage{cleveref} 
\usepackage{tabularx}
\usepackage{epstopdf}
\usepackage{comment}
\definecolor{mycolor1}{RGB}{0,176,80}
\definecolor{mycolor2}{RGB}{112,48,160}

\def\BibTeX{{\rm B\kern-.05em{\sc i\kern-.025em b}\kern-.08em
    T\kern-.1667em\lower.7ex\hbox{E}\kern-.125emX}}
\begin{document}

\title{Compact Visual Data Representation for Green Multimedia - A Human Visual System Perspective}
\author{Peilin Chen, 
        Xiaohan Fang, 
        Meng Wang, 
        Shiqi Wang, \textit{Senior Member, IEEE}, 
        and Siwei Ma, \textit{Fellow, IEEE}
        
    \thanks{
    P. Chen, X. Fang, and S. Wang are with the Department of Computer Science, City University of Hong Kong, Hong Kong SAR, China (e-mail: plchen3@cityu.edu.hk; xhfang3-c@my.cityu.edu.hk; shiqwang@cityu.edu.hk).  

    M. Wang is with the School of Data Science, Lingnan University, Hong Kong SAR, China. (e-mail: mengwang7@ln.edu.hk).

    S. Ma is with the National Engineering Research Center of Visual Technology, School of Computer Science, Peking University, Beijing 100871, China  (e-mail: swma@pku.edu.cn).
    
    Corresponding author: Shiqi Wang.
    }
}

\maketitle

\begin{abstract}
The Human Visual System (HVS), with its intricate sophistication, is capable of achieving ultra-compact information compression for visual signals. 
This remarkable ability is coupled with high generalization capability and energy efficiency. By contrast, the state-of-the-art Versatile Video Coding (VVC) standard achieves a compression ratio of around 1,000 times for raw visual data. This notable disparity motivates the research community to draw inspiration to effectively handle the immense volume of visual data in a green way. 
Therefore, this paper provides a survey of how visual data can be efficiently represented for green multimedia, in particular when the ultimate task is knowledge extraction instead of visual signal reconstruction. 
We introduce recent research efforts that promote green, sustainable, and efficient multimedia in this field. 
Moreover, we discuss how the deep understanding of the HVS can benefit the research community, and envision the development of future green multimedia technologies.

\end{abstract}


\section{Introduction}
In the digital era, the proliferation of visual data has surged to unprecedented levels, fueled by the widespread use of visual data acquisition sensors~\cite{chen2020internet} and the rise of artificial intelligence-generated content (AIGC)~\cite{foo2023ai}. The booming presents significant challenges in ``green'', in terms of efficiently storing, transmitting, and analyzing visual information at low cost. 
\textcolor{black}{
For example, high-resolution video streaming, which accounts for the majority of internet traffic, necessitates substantial computational resources and bandwidth, leading to a significant carbon footprint. The traditional infrastructure, relying heavily on centralized data centers, further exacerbates this impact due to energy-intensive cooling requirements and long-distance data transmission. 
According to Cisco's Annual Internet Report, video streaming and downloads constituted over 82\% of all internet traffic in 2022, emphasizing the need for more efficient visual data handling mechanisms.
}

\textcolor{black}{
In general, green multimedia encompasses a series of innovative strategies aimed at reducing the carbon footprint of multimedia handling, which includes diverse aspects ranging from data storage, processing, and transmission. This paper explores three critical dimensions of green multimedia technologies that could significantly enhance energy efficiency across various multimedia platforms. 
\textit{\textbf{Green Storage through Compression}}: The first aspect focuses on the storage of multimedia content. By employing advanced compression techniques, it is possible to drastically reduce the data size, thereby lowering the storage requirements and associated energy costs, and enabling more efficient data management. Although video compression standards such as H.264/AVC~\cite{wiegand2003overview}, H.265/HEVC~\cite{sullivan2012overview}, as well as H.266/VVC~\cite{bross2021overview}, AV1~\cite{chen2018overview} and AVS3~\cite{8954503} have considerably reduced data sizes at the same quality level, their theoretical limits of compression efficiency~\cite{shannon1948mathematical} is being constantly approached. This highlights the urgent need for innovative breakthroughs for compact visual data representation for ``green''.
\textit{\textbf{Green Processing by Direct Feature Usage}}: 
The second dimension examines the processing methodologies within multimedia frameworks. Traditional approaches often involve decoding and then processing to extract features from data for downstream usage, which can be computationally expensive due to the operations in these two stages. However, by integrating processing capabilities directly at the decoder level and eliminating the need for separate decoding steps, significant energy reductions can be achieved. 
For example, the feature stream in the Digital Retina~\cite{gao2021digital} can enable the receiver to directly use the received features in object detection, which achieves comparable task performances with several hundredfold lower bandwidth as well as fewer inference operations.
This strategy can enhance processing efficiency and reduce the energy overhead of multimedia systems, facilitating greenness.
\textit{\textbf{Dynamic Green Operations}}: 
The third facet surveys the dynamic aspects of multimedia data handling. By enabling end systems to take full advantage of available resources or use only the necessary data needed for specific tasks adaptively, it ensures that no superfluous data processing occurs along the communication chain. This adaptability not only minimizes unnecessary energy expenditure but also enhances the overall responsiveness and efficiency of multimedia services. 
For instance, the scalable coding framework~\cite{zhang2023rethinking} suggests using a modular bitstream based on network conditions. This approach reduces encoder operations by skipping bitstream construction for the enhancement layer when bandwidth is low.
Nonetheless, the decoder can still reconstruct results through adaptive operations, which meets the envision of green multimedia.
Furthermore, according to recent studies in Green Metadata~\cite{herglotz2023extended, herglotz2024energy, herglotz2024complexity}, the reduction of decoding complexity could vastly decrease energy savings.
}

\begin{figure*}
    \centering
    \includegraphics[width=0.8\linewidth]{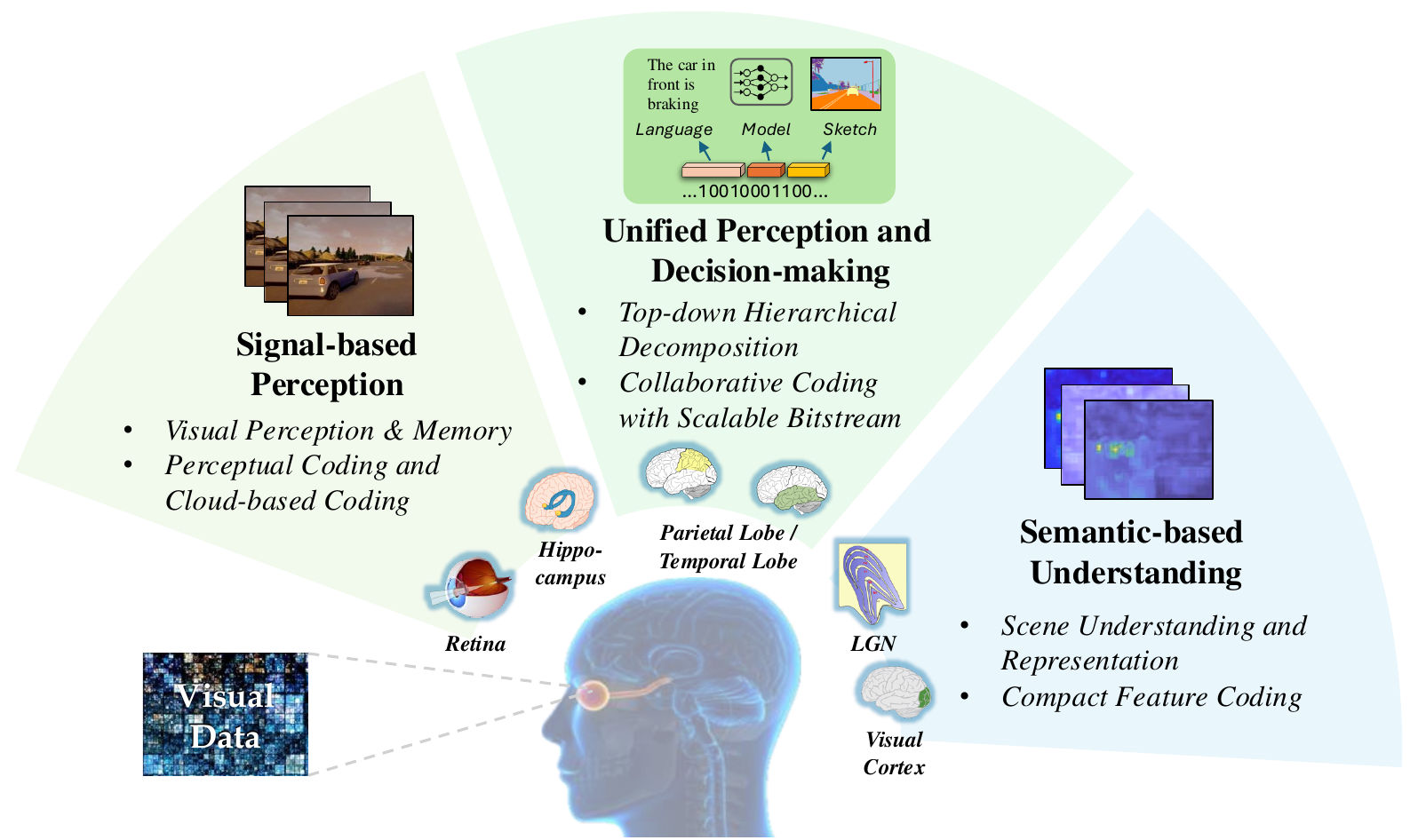}
    \caption{This diagram depicts how the psychophysical HVS features can be utilized for compact visual data representation. }
    \label{fig:hvs_inspired}
\end{figure*}

When advancing green multimedia technologies, the Human Visual System (HVS)~\cite{kruger2012deep} presents an intriguing benchmark in this context, which can achieve an astounding compression efficiency, and compact visual information around 100,000 times while maintaining high generalization and energy efficiency~\cite{brain_report}.
It vastly outperforms the current Versatile Video Coding (VVC) standard, which manages a compression ratio of only about a thousand times for raw visual data. 
Drawing inspiration from the HVS's capabilities in artificial systems for visual data representation marks a promising research direction~\cite{zhang2020system}.

Despite the existing technological gaps, considerable efforts are underway to incorporate novel ideas from the HVS to advance visual data compression. Noteworthy developments such as perceptual video coding,  Video Coding for Machines (VCM)~\cite{duan2020video} and ultra-low bitrate generative compression~\cite{chen2024generative} illustrate strides toward significantly minimizing data sizes while preserving essential information for perceptual reconstruction, downstream analysis, and semantic restoration. These innovations indicate a shift toward more compact and energy-efficient multimedia representation beyond the traditional signal-based visual data coding, aligning with the broader goal of green technology development.

This paper provides a survey of the current landscape in visual data representation aiming at green multimedia, emphasizing how HVS can motivate compact video representation, in particular perceptual coding, compact feature representation, and collaborative representation. 
We also delve into recent efforts to transcend the limitations of traditional hybrid video coding frameworks, spotlighting cutting-edge strategies that promise ultra-low bitrate representation to achieve greenness in the bitstream.
Moreover, we discuss how these advancements influence the efficiency and sustainability of multimedia analytics frameworks. 
Through this review, we aim to provide a comprehensive understanding of how advancements in compact data representation can foster greener multimedia, which underscores the ongoing importance of research and innovation in achieving optimal compression efficiency alongside environmental sustainability for managing visual data.

\section{Exploring the Human Visual System: Insights and Implications for Green Multimedia}

\textcolor{black}{
As we strive toward environmentally sustainable multimedia systems, understanding the principles underlying the HVS provides invaluable insights into developing energy-efficient and resource-conscious technologies for green multimedia. The HVS exemplifies a system evolved to interpret vast amounts of visual information with remarkable efficiency, offering a model for multimedia processing that balances high-quality perception with low resource usage~\cite{wang2006modern, kruger2012deep}. By emulating these efficiencies, green multimedia technologies can minimize bandwidth, storage, and computational requirements, thereby reducing their environmental footprint.
}
This system's ability to discern minute details, recognize patterns, and process complex scenes in real-time offers invaluable insights for developing green multimedia systems.
In particular, the HVS is adept at transforming light into electrical signals through intricate biochemical organizations in the retina, primarily via rods and cones. 
These signals are then refined and relayed through various cortical areas, each specializing in different aspects of vision, such as color, spatial frequency, and motion perception~\cite{remington2011clinical}. 
This process highlights a hierarchical and highly efficient method of visual data processing, from basic sensory input to complex perceptual interpretations.
Central to this efficiency is the HVS's ability to operate at extremely low bitrate representations, particularly from the primary visual cortex (V1) to extrastriate cortical areas~\cite{sziklai1956some}. 

\begin{figure*}
    \centering
    \includegraphics[width=0.98\linewidth]{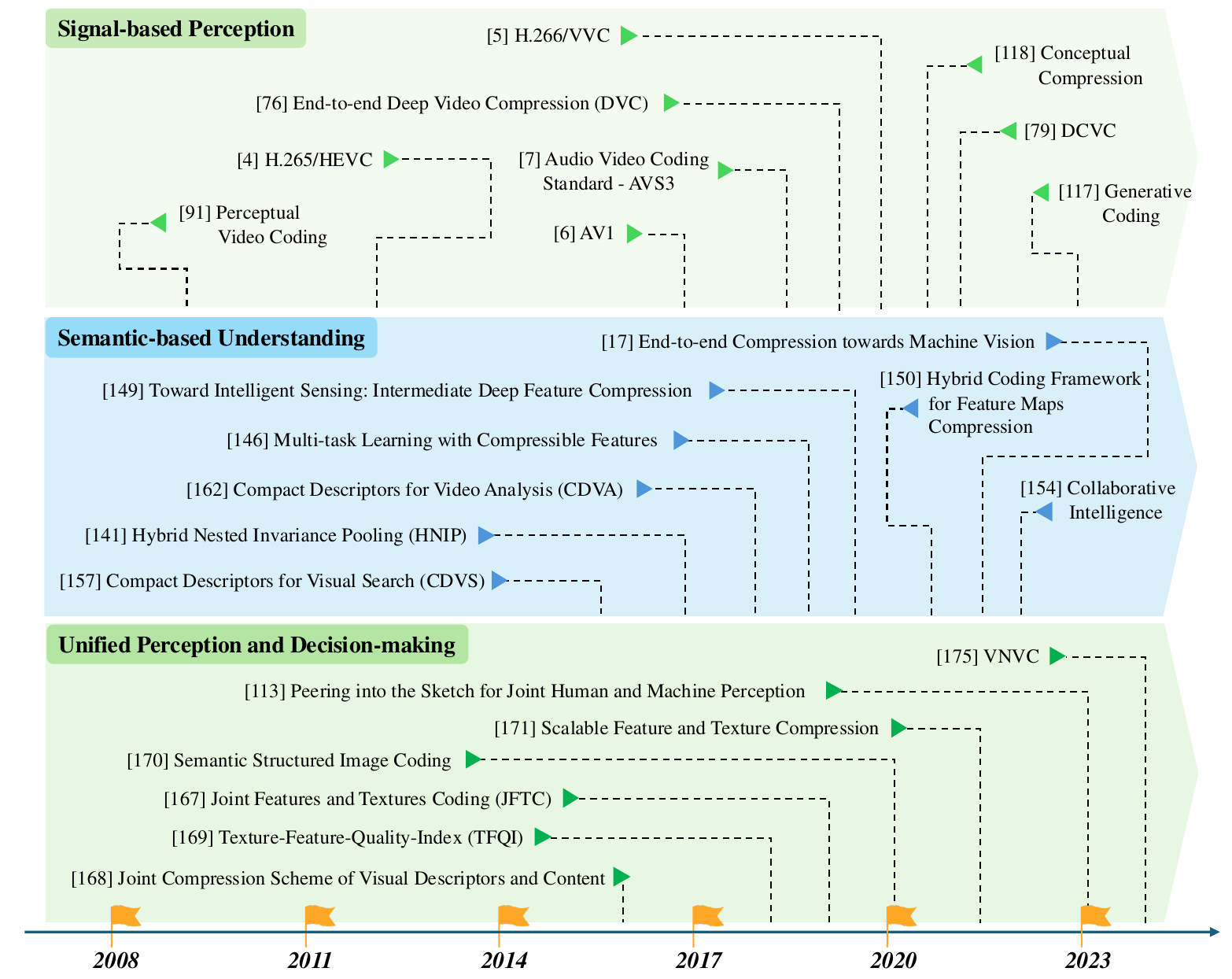}
    \caption{The roadmap of compact visual data technologies regarding signal-based perception, semantic-based understanding, as well as unified perception and decision-making. }
    \label{fig:roadmap}
\end{figure*}

Therefore, there are three distinguished properties of HVS that can be leveraged in the design of compression algorithms toward green multimedia, as shown in Fig. \ref{fig:hvs_inspired}. 
First, the non-uniform distributions of cone photoreceptors and ganglion cells, the varying sensitivity with respect to spatial/temporal frequency, and the adaptation to lighting motivate the unique characteristics of the visual system that are used to develop perceptual coding algorithms. 
In addition, the memory mechanism of HVS also inspires effective compression algorithms based on cloud/external resources. 
Second, the task-driven feature extraction based on the brain's selective attention and processing of visual information motivates the paradigm of ``visual coding for machine''. 
In particular, the features instead of textures play the central role, and compact feature representation technologies have been subsequently developed. 
Third, neural communication that occurs within the layers of the visual cortex motivates the collaborative and scalable coding approach that adapts to different purposes in utilizing the visual information. 
Such capabilities inspire the design of systems that not only mimic these biological efficiencies but also adapt them for digital multimedia purposes.

In this context, the ``Digital Retina" as a transformative framework~\cite{lou2019towards,lou2020towards,gao2021digital}  envelopes most of the characteristics of HVS in designing green multimedia systems. 
This novel architecture emphasizes a dual representation strategy: signal-based compact textures for essential visual information retention and semantic-based compact features crucial for machine vision applications. 
The signal component ensures that necessary details for human perception and comprehensive analytics are preserved, while the semantic component abstracts higher-level features into more compact forms, significantly reducing data size and enhancing analytical processes~\cite{gao2021digital}. 
The Digital Retina also introduces a dynamic adaptation mechanism through model parameter updating, where deployed models can continuously refine their data interpretation capabilities. 
This flexibility can extend to its transmission protocols, which adjust data throughput based on current network conditions and analytic demands, optimizing overall resource usage and ensuring efficient real-time data processing. 
Moreover, by focusing on transmitting abstracted feature-level data, the Digital Retina can take full advantage of the front-end intelligence while maintaining user privacy. 

While achieving ultra-compact and energy-efficient processing capabilities comparable with the HVS remains a distant goal, numerous research initiatives are actively pushing the boundaries of compact visual data representation.
These works are steadily advancing the field, providing innovative solutions that bring us closer to the efficiency and effectiveness of biological systems. 
The following sections of this paper will delve into these developments, focusing on the three critical areas of compact video compression, compact feature representation, and collaborative representation, each of which plays a vital role in the ongoing quest for greener multimedia systems. In particular, a roadmap was provided as Fig. \ref{fig:roadmap}. Finally, we will discuss the connections of these technologies with the recent advances in edge-cloud computing, AIGC models, and large vision language models. 

\section{Compact Visual Data Representation}

\textcolor{black}{
As multimedia usage expands, the demand for efficient data representation has intensified, especially in video-centric applications that consume substantial bandwidth and energy resources. In the context of green multimedia, compact visual data representation addresses the critical need to reduce carbon footprint, energy consumption, and resource usage without compromising the quality and usability of visual data. This section explores various techniques and technologies focused on compressing and efficiently representing visual data, each contributing uniquely to the objectives of green multimedia.
}
\subsection{Compact Video Compression}
\textcolor{black}{
Efficient video compression is paramount for green multimedia, as it reduces data volume, storage demands, and transmission energy. Traditional video compression standards and end-to-end coding technologies aim to minimize the bitstream size while preserving visual fidelity, directly reducing the bandwidth and storage required for video applications. Perceptual compression techniques optimize data representation by focusing on the parts of visual content most relevant to human perception. 
By simulating how HVS perceives the information, these methods achieve a more efficient data size without impacting user experience. 
In the green multimedia context, perceptual compression techniques enhance energy efficiency by prioritizing the processing, storage, and transmission of perceptually critical data, which can be implemented via HVS-guided rate-distortion optimization.
Compression based on external data utilizes additional information, such as prior models or reference datasets, to facilitate higher compression efficiency. This approach enables a substantial reduction in bit rate, lowering both storage and transmission energy demands. Within green multimedia, external data compression offers a promising route for sustainable data handling by minimizing resource requirements.
}

\subsubsection{Video Coding Standards}

The widely applied hybrid video coding framework strategically combines prediction, transformation, quantization, and entropy coding modules, which collectively work to eliminate diverse redundancies across spatial, temporal, perceptual, and statistical dimensions. 
While traditional coding technologies may not be the central theme of this paper, they form the foundational basis, and many subsequent methods reflect their underlying design principles. 
\textcolor{black}{
In particular, Versatile Video Coding (VVC)~\cite{bross2021overview}, or H.266, is the latest standard from the MPEG that offers significant compression improvements, potentially halving data needs compared to HEVC while maintaining video quality, ideal for 8K and VR applications. AOMedia Video 1 (AV1)~\cite{chen2018overview,eichermuller2024svt} is an open, royalty-free codec designed for internet video, outperforming HEVC in efficiency and supported by major industry players for improved performance in streaming scenarios. AVS3~\cite{8954503}, from China's Audio Video Coding Standard Workgroup, targets ultra-high-definition content with high compression and robust error recovery, addressing the vast data demands of network transmission and online services in China and beyond.
}
Additionally, as digital devices proliferate and the volume of visual data swells, the state-of-the-art video coding standards such as VVC and AVS3 have adopted advanced coding tools, including the more flexible partitioning schemes~\cite{8712764,8917942} with quad-tree plus binary-tree and ternary tree structures, alongside advanced intra and inter prediction techniques~\cite{9115218}. 

More specifically, 67 modes have been designed for accommodating diverse video content textures~\cite{Modes67}. 
This granularity enhances prediction accuracy and reduces residual signaling burdens. 
Meanwhile, the adoption of multiple reference lines and the Cross-Component Linear Model (CCLM) predictions further leverage inter-channel redundancies for chroma intra coding~\cite{zhang2018enhanced, 9105802, 9105965}, showcasing significant efficiency in handling color components. 
Moreover, the VVC and AVS3 standards incorporate a range of sophisticated intra prediction tools tailored for processing non-natural video content. 
These include intra block copy~\cite{IBC_AVS3,HBVP}, palette mode~\cite{PLT}, and string copy prediction~\cite{StringCopy}. 
On the inter prediction front, the standards excel at reducing temporal redundancies. 
To handle the complex motions in real-world scenes, VVC and AVS3 have integrated affine motion compensation techniques~\cite{affine_avs, affine_tip}, designed to handle non-translational movements like zooming and rotation. 
The rise in video resolution further complicates MV signaling, prompting the development of sophisticated MV coding and refinement strategies. These include adaptive motion vector resolution (AMVR)\cite{amvr}, history-based motion vector prediction (HMVP)\cite{HMVP_DCC}, advanced motion vector prediction (AMVP)\cite{mccann2010samsung}, bi-directional optical flow (BDOF)\cite{alshin2010bi}, merge mode with motion vector difference (MMVD)\cite{jeong2020merge}, and decoder-side motion vector refinement (DMVR)\cite{DMVR_CSVT}. 
For residual representation in both intra and inter prediction, trellis-coded quantization (TCQ)\cite{Trellis_DCC} is employed, where coefficients within a coding block are mapped onto a trellis graph to determine the path with the lowest rate-distortion cost, effectively minimizing redundancies. 
These standards support multiple discrete cosine or sine-based transform (DCT/DST) cores, which help to more efficiently concentrate residual energies in VVC\cite{TransformVVC} and AVS3~\cite{IST_ICMEW}. 
Historically, overlapped block motion compensation has been used to enhance the accuracy of inter prediction~\cite{orchard1994overlapped}. 
Finally, the elimination of statistical redundancy is achieved through context-adaptive binary arithmetic coding (CABAC)~\cite{HEVC_CABAC}, which merges adaptive binary arithmetic coding with context modeling for statistical redundancy removal.
\textcolor{black}{
Regarding encoding optimization, a multi-grained parallel solution (MPS) was proposed to optimize HEVC encoding on heterogeneous systems by leveraging massively parallel motion estimation on GPUs and adaptive parallel mode decisions on CPUs, demonstrating significant improvements in concurrent execution efficiency~\cite{8713491}.
}
\textcolor{black}{
Recently, the Enhanced Compression Model (ECM), a further enhancement of the VVC, has undergone extensive studies with a variety of advanced coding tools. These tools encompass various coding modules such as intra/inter prediction, aiming to further enhance its performance~\cite{youvalari2023filtered,abdoli2024video,pu2024intra,liao2024improvement,deng2024inter, vishwanath2023discrete}.
}

\textcolor{black}{
In addition to traditional two-dimensional (2-D) videos, a variety of three-dimensional (3-D) visual representations is becoming increasingly popular for creating vivid, immersive virtual user experiences, which facilitates the development of volumetric videos.
These representations include point clouds~\cite{berger2017survey, schwarz2018emerging}, meshes~\cite{maglo20153d}, 360-degree VR videos~\cite{zhou2019video}, Neural Radiance Fields~\cite{mildenhall2021nerf}, and 3D Gaussian Splatting~\cite{kerbl20233d}. 
Although volumetric video offers an immersive experience that surpasses traditional 2-D video, it commonly comes with the challenge of large raw data sizes. 
To address this, specialized visual data compression methods are essential to compact them and manage energy consumption, particularly in applications that require low-latency interactions and efficient storage solutions, which facilitate the achievement of green multimedia.
Efficiently compressing volumetric video involves various innovative techniques. One strategy is reshaping the data into 2D-frame-like video sequences, which can then be processed using established video coding schemes. For instance, light field data can be decomposed by view direction and arranged into inter-correlated sequences for encoding with VVC standards. Similarly, the Video-based Point Cloud Compression (V-PCC) technique developed by the MPEG projects 3-D point clouds onto 2-D maps, allowing existing 2-D video codecs to compress the data effectively, with remapping back to 3-D conducted at the decoder~\cite{jang2019video, li2020efficient, li2019advanced, akhtar2021video}. Another method for compact volumetric video compression utilizes the geometry characteristics from 3-D data, such as the Geometry-based Point Cloud Compression (G-PCC) standard, which employs a pruned octree format to approximate the original data efficiently~\cite{graziosi2020overview, ainala2016improved}. To enhance the efficiency of MPEG Immersive Video (MIV) coding, researchers have proposed a method~\cite{dziembowski2023immersive} based on modifying patch average color and adjusting the dynamic range of depth atlases.
}

\subsubsection{End-to-end Video Coding Technologies}

The end-to-end (E2E) coding framework serves as an alternative solution for compression. Unlike traditional methods that often involve separate components for different stages of compression, E2E frameworks integrate all coding processes from input to output into a unified system. This integration allows the system to learn and optimize the compression pipeline holistically, using deep learning techniques to adaptively manage data based on content complexity and desired fidelity levels. By training these systems directly on the target data, they can potentially achieve higher efficiency and better quality compared to modular approaches.
In exploring end-to-end visual data compression, a notable pioneer contribution is made by Ballé \textit{et al.}, who designed an end-to-end architecture that includes a decorrelation layer known as Generalized Divisive Normalization (GDN), and replaced traditional step quantization with uniform noise addition during training to highlight the advantages of end-to-end methods \cite{balle2016end, balle2016end2}. 
Subsequent studies have typically utilized the upper bound entropy of latent codes as a surrogate for the rate term in rate-distortion optimization. 
Ballé \textit{et al.} enhanced their architecture with a hyper-prior network that aids in decoding by leveraging side information extracted from latent codes, presuming a Gaussian distribution \cite{balle2018variational}. 
This approach was further refined by Cheng \textit{et al.}, who employed a Gaussian mixture model instead of a simple Gaussian for a more precise entropy estimation \cite{cheng2020learned}. 
Chang \textit{et al.} proposed a framework that utilizes deep structural and textural information to enhance compression versatility \cite{chang2020conceptual}.
\textcolor{black}{
Tang \textit{et al.} proposed an end-to-end image compression method that integrates graph attention and asymmetric convolutional neural networks (ACNN) to enhance local key feature effects, reduce training costs, and achieve state-of-the-art rate-distortion performance~\cite{9858899}.
}
\textcolor{black}{
Recently, a hybrid video compression framework based on the ECM~\cite{zhao2024neural} has been proposed to integrate deep learning techniques, achieving significant BD-rate savings for Y, U, and V components under random access configuration.
}
The field of end-to-end video compression has also seen significant advancements. 
Lu \textit{et al.} developed deep video compression (DVC), which is the first comprehensive model for video compression that optimizes nearly all aspects of the framework, including the use of a hyper-prior model for intra-frame compression and an optical flow network for effective motion estimation and compensation \cite{lu2019dvc}. 
They later enhanced this model with an adaptive quantization layer to support variable bitrate coding \cite{lu2020end}. 
Building on traditional video coding techniques, Yang \textit{et al.} introduced a strategy that incorporates hierarchical quality sub-models, yielding results competitive with the fastest modes of the HEVC standard \cite{yang2020learning}. 
Recently, Li \textit{et al.} introduced a deep contextual video compression method named DCVC~\cite{li2021deep}, which shifts from the traditional residual coding strategy to a conditional coding framework. 
Instead of using subtraction to eliminate temporal redundancy, DCVC treats temporal prediction as conditions, enabling the codec to mine the potential dependencies implicitly. 
Subsequently, Sheng \textit{et al.} introduced DCVC-TCM~\cite{sheng2022temporal}, which enhances DCVC by integrating multi-scale temporal predictions throughout the encoding process and using them for entropy modeling. 
Li \textit{et al.} expanded on this with DCVC-HEM~\cite{li2022hybrid}, adding a latent prior to further optimize temporal correlation. 
By increasing the diversity of contexts in DCVC, DCVC-DC was proposed to achieve superior coding performance compared with VVC~\cite{li2023neural}. 
\textcolor{black}{
Recently, DCVC-FM~\cite{li2024neural} adopted feature modulation to support a wider quality range and maintain performance under long prediction chains, achieving significant bitrate savings and computational efficiency improvements.
NVC-1B~\cite{sheng2024nvc} is the first neural video coding model with over 1 billion parameters, demonstrating significant improvements in video compression performance and setting a new state-of-the-art in compression efficiency.
}
\textcolor{black}{
While neural network-based compression techniques can achieve high performance in terms of compression efficiency and subjective quality, their computational overhead and energy consumption still raise important sustainability concerns. Reducing this complexity without sacrificing compression effectiveness represents a critical challenge in green multimedia research. Current strategies address this trade-off by exploring model compactness techniques, such as pruning and quantization~\cite{guo-hua2023evc}, which can dramatically lower computational requirements, and by leveraging knowledge distillation to transfer learned features into more lightweight models~\cite{10.1145/3650034}. Beyond these approaches, innovative transforms and hybrid architectures that fuse traditional coding techniques with learned components offer the potential to preserve much of the efficiency gain from neural models while limiting their energy footprint~\cite{10.1145/3652148}. As the field continues to mature, the development of even more sophisticated complexity-control mechanisms—supported by new theoretical insights and hardware acceleration strategies—will be essential. By prioritizing methods that harmonize compression efficiency with low energy consumption, future research can help ensure that advanced compression technologies remain both environmentally responsible and technically competitive.
}


\subsubsection{Perceptual Compression}
\textcolor{black}{
In the context of green multimedia, perceptual coding offers an energy-efficient approach by leveraging the HVS to focus computational resources on areas that significantly impact perceived quality. \textcolor{black}{By reducing the processing complexity in regions less critical to visual experience, perceptual coding techniques can decrease the amount of data required for transmission and storage, thus reducing bandwidth and energy consumption across the multimedia pipeline without compromising user experience~\cite{kannur2009power,hemmati2013game,mazumdar2019perceptual}}. 
}
Perceptual compression is a long-standing research topic that has attracted numerous attentions~\cite{wu2013perceptual, zhang2023survey}. 
The fundamental principle is to optimize compression efficiency by simulating visual perception and exploiting visual redundancies. 
\textcolor{black}{
While PSNR remains a widely used metric due to its simplicity and friendly hardware implementations, various visual sensitivity models have been developed to better align with perceptual quality requirements.
}
Hence, many visual sensitivity models have been proposed by studying the characteristics of visual distortions. 
Specifically, the feature-similarity (FSIM) index has been introduced, which considers the contribution of low-level features perceived by HVS~\cite{zhang2011fsim}.
The nature of compression artifacts that affect HVS also necessitates research to characterize and mitigate their impact on perceptual quality~\cite{zeng2014characterizing}. 
\textcolor{black}{The Video Multimethod Assessment Fusion (VMAF) metric~\cite{vmaf} offers a highly efficient full-reference video quality assessment method suitable for a wide range of video coding scenarios.}
\textcolor{black}{
Specifically, Luo \textit{et al.} have designed a VMAF oriented perceptual coding Based on piecewise metric coupling, which achieves promising bit savings while better aligned with HVS~\cite{luo2021vmaf}.
}
Recent advancements reveal that deep features from neural networks can significantly enhance perceptual similarity assessments, proposing a paradigm shift in how perceptual quality is evaluated in compression~\cite{zhang2018unreasonable}.
The utilization of visual Just Noticeable Difference (JND) modeling is also one of the research directions of perceptual compression~\cite{yang2005motion, ki2018learning, tian2020just, nami2022bl}. These approaches leverage the fact that human eyes cannot detect changes below the JND threshold around a pixel, due to their inherent spatial and temporal sensitivity as well as the masking properties~\cite{241504}.
Visual Attention (VA), which is also referred to as saliency, is a cognitive process that directs the eye's focus toward Regions of Interest (ROI) to capture finer details. 
This mechanism can also be leveraged to effectively advance the perceptual compression~\cite{hadizadeh2013saliency, wang2011ssim,mazumdar2019perceptual}.
\textcolor{black}{In particular, Mazumdar \textit{et al.} proposed a saliency-based perceptual compression technique named Vignette, which can cooperate with off-the-shelf compression codecs to improve cloud storage capacity and reduce power consumption in mobile devices~\cite{mazumdar2019perceptual}.}
Furthermore, Chen \textit{et al.} introduced a novel approach with the proxy network to optimize the end-to-end compression model against quantitative perceptual performance~\cite{chen2020proxiqa}.
Rouis \textit{et al.} adapts the Lagrangian multiplier based on perceptual criteria within the HEVC standard, aiming to enhance coding efficiency by focusing on visually significant areas of the video~\cite{rouis2018perceptually}.
\textcolor{black}{
For perceptually compressed VR videos, the video decoder architecture named mVDO was proposed, which maps peripheral video regions to decoder cores in multiple frequency domains to achieve better energy efficiency~\cite{mazumdar2020perceptual}.
}
\textcolor{black}{
The perceptually tuned ECM has also been studied~\cite{philippe2024ed}, which aims to maximise the perceived quality.
}
Additionally, spatial domain pre-processing is a popular strategy where blur intensity is tailored to make less significant regions more easily encodable.
To make preprocessing compatible with diverse codecs, Chadha \textit{et al.} proposed a rate-aware perceptual-oriented framework comprised of dedicated perceptual loss and rate loss to achieve single-pass operation before video coding~\cite{chadha2021deep}.
\textcolor{black}{
Arai \textit{et al.}~\cite{arai2024gop} proposes a novel GOP-based method that enhances rate-distortion performance in terms of PSNR and other quality metrics by jointly optimizing a preprocessing model with the video compression model.
}
One of the primary benefits of these preprocessing techniques is their compatibility with any standard codec, eliminating the need for modifications. 
Nonetheless, since the quality control and encoding processes are independent, the gains from preprocessing are generally considered a lower bound on the potential improvements in perceptual compression. 
Moreover, combining visual encoding with existing standards allows for seamless integration into current systems without additional hardware or decoder updates. 
In general, video coding standards, including H.264/AVC, AVS2, HEVC, AVS3, and VVC, support flexible encoding parameters across different regions, enabling the assignment of varied quality parameters in a frame. This approach allows for finer quality distinctions, with critical visual areas encoded at higher quality due to lower QPs.
\textcolor{black}{
Furthermore, exploring non-PSNR distortion measures, such as SSIM or VMAF, in conjunction with ROI techniques presents a promising avenue for future research, particularly in the context of power saving. While these methods are known to enhance visual quality, their potential for power reduction remains underexplored and warrants further investigation. Future researchs can explore the relationship between perceptual video coding and power efficiency, examining how these advanced distortion measures can contribute to energy savings at various video processing stages.
}

\subsubsection{Compression based on External Data}
Apart from the compression methods that aim to leverage the original signal statistics for redundancy removal, there are also compact representation extraction methods based on cloud or external resources, which utilize the memory scheme of HVS for external-based coding or priors-guided generative compression. 
As a pioneer work, Yue \textit{et al.} proposed to leverage a large-scale image database to compress images based on their down-sampled versions and local feature descriptors~\cite{yue2013cloud}. This approach uses descriptors to search for similar images in the cloud, identifies matching patches, and stitches them to reconstruct high-quality images.
To take full advantage of existing video data in the cloud, Wang \textit{et al.} proposed a graph-based clustering approach and corresponding compression framework to compactly represent the abundant near-duplicate video data~\cite{wang2016joint}. 
\textcolor{black}{
Recently, it has also been shown that external resources could provide abundant guidance during the deep reconstruction in decoding, which facilitates both promising machine analysis and human perception performance under ultra-low bitrate conditions~\cite{mao2023peering}.
}

In general, generative coding approaches can also be regarded as a type of compression based on external data, because they commonly model pre-collected data characteristics and extract useful priors that could be treated as reconstruction context in the decoding.
Specifically, Oquab \textit{et al.} proposed using generative models for the chat video compression, which adopts the facial landmarks as the compact representation and generative priors, achieving noticeable bandwidth saving \cite{oquab2021low}. 
Tang \textit{et al.} studied the face key point as the compact representation for efficiently denoting the non-key frames in a generative compression framework, which not only maintains reconstructed signal fidelity but also achieves ultra-low bitrate communication~\cite{tang2022generative}. 
Li \textit{et al.} proposed utilizing the pre-defined 3-D descriptors to compactly model the appearance representation, which can achieve over 50\% reduction in coding bits compared to VVC under ultra-low bitrate conditions~\cite{li2022towards}. 
To comprehensively model the complex dynamics between frames and construct ultra-compact bitstream for generative face reconstruction, Chen \textit{et al.} implicitly projects motion to high-dimensional space as the intermediate data, which largely reduces the bandwidth requirement while providing a vivid animation performance with the decoding frame~\cite{chen2023compact}.
Chang \emph{et al.} proposed a novel conceptual compression approach, utilizing a dual-layered model with structure and texture components encoded separately and synthesized via a hierarchical fusion GAN (HF-GAN) for high-quality visual reconstruction, showcasing enhanced versatility in bitrate efficiency and content manipulation~\cite{chang2022conceptual}.
Wang \emph{et al.} introduced a dynamic multi-reference prediction method for ultra-low bitrate generative face video compression that effectively handles large head motions. By utilizing key maps and multiple dense motion maps from current and diverse references, the method dynamically refreshes the reference frame to minimize motion discrepancies~\cite{9897729}.
\textcolor{black}{
Zhang \emph{et al.} introduced a unified Cross-Modality Video Coding (CMVC) paradigm that leverages Multimodal Large Language Models (MLLMs) for video compression. This approach disentangles video into spatial and motion components, optimizing video reconstruction through distinct encoding-decoding modes, such as Text-Text-to-Video (TT2V) and Image-Text-to-Video (IT2V), with experimental results demonstrating superior semantic reconstruction and perceptual consistency~\cite{zhang2023rethinking, zhang2024video}.
}

\subsection{Compact Feature Compression}
\textcolor{black}{
Beyond raw visual data, feature-based compression focuses on reducing the size of extracted features rather than full video frames. This technique enables efficient downstream tasks such as object recognition or motion analysis directly from compressed feature representations, bypassing the need for full data reconstruction. In green multimedia, feature compression contributes by reducing the computational overhead of decoding entire frames, thus saving energy and reducing latency.
}
Specifically, enabling the transmission of feature stream is the most unique characteristic of digital retina~\cite{gao2021digital}. It is fundamentally inspired by the fact that when facing dynamic vision tasks, the human brain has a unique feedback mechanism that fulfills different demands and purposes~\cite{zhang2020system} by extracting the corresponding features. 
The development of compact feature descriptors is driven by the requirements of front-end intelligence, where features extracted at very low bit rates from edge devices can trigger simple downstream tasks such as object search queries. 
This approach follows the ``analyze-then-compress" paradigm, which contrasts with the traditional ``compress-then-analyze" method, where images or videos are first compressed at the signal level~\cite{redondi2016compress}. 
Specifically, the well-known handcrafted descriptor, Scale Invariant Feature Transform (SIFT), can undergo compression through methods like spectral hashing~\cite{weiss2008spectral}, transform coding~\cite{chandrasekhar2009transform}, and product quantization~\cite{jegou2010product, ge2013optimized}. 
Furthermore, to meet the demands for high-speed Hamming distance comparisons, binary descriptors such as Binary Robust Independent Elementary Features (BRIEF)\cite{calonder2011brief}, Oriented FAST, and Rotated BRIEF (ORB)\cite{rublee2011orb}, Binary Robust Invariant Scalable Keypoints (BRISK)\cite{leutenegger2011brisk}, and UltraShort Binary Descriptor (USB)\cite{zhang2014usb} have been introduced. 
For global descriptor compression, which integrates local descriptors, approaches such as Vector of Locally Aggregated Descriptors (VLAD)\cite{jegou2010aggregating} and Fisher Vectors (FV)\cite{perronnin2010large} have been extensively studied~\cite{chen2013residual,chen2011residual,lin2014rate}. 
With advancements in learning-based methods, techniques such as Kernel-based Supervised Hashing (KSH)\cite{liu2012supervised} have emerged to provide concise yet discriminative representations using kernel-based models. 
A similar supervised hashing framework was also introduced using regularization techniques\cite{shen2015supervised}. 
\textcolor{black}{
More recently, deep learning-based hashing methods~\cite{luo2023survey} have been developed to deliver end-to-end hashing representations for more efficient retrieval, including center-based binary neural network optimization~\cite{he2024one}, deep product quantization~\cite{gu2024entropy} and cross-modal hashing techniques~\cite{tu2024two}. 
}

Recently, deep learning methods have made substantial contributions to a variety of computer vision applications via extracted features, marking significant progress across numerous domains through their data-driven capabilities. 
These features offer robust representational strength and reduce the dimensionality of the pixel space in the original visual data. 
The introduction of Hybrid Nested Invariance Pooling (HNIP) has further enhanced feature representation by encoding translation, scale, and rotation invariances directly into the features \cite{lin2017hnip}. Building on these concepts, new methods in video coding have been adapted for deep feature compression, defining three distinct feature categories: independently-coded, predictively-coded, and skip-coded features \cite{ding2017rate}. To minimize redundancy, both local and global features are combined for more effective visual searches \cite{ding2020joint}. This approach leads into the broader idea of collaborative intelligence, where computational tasks are divided between edge devices and the cloud, optimizing both inference and feature handling \cite{kang2017neurosurgeon}. Research in feature compression, feature completion, and bit allocation has expanded, as evidenced by various studies \cite{bragilevsky2020tensor,alvar2019multi,alvar2020bit,cohen2020lightweight}. Additionally, Chen \textit{et al.} have systematically explored the compression of intermediate features for intelligent sensing applications \cite{chen2019toward}, and have proposed methods such as repacking modes and pre-quantization to enhance data fidelity \cite{chen2020data}. Based on the concept of end-to-end visual data compression frameworks, Wang \textit{et al.} have conducted an extensive study on neural network architectures and optimization objectives to achieve compact semantic representations of features~\cite{wang2021end}. A unified optimization framework was subsequently proposed to enhance the end-to-end image compression method for more efficient task-oriented feature compression, which sheds on how the compact feature construction can take advantage of existing visual data compression methods.
In a similar vein, a novel end-to-end learned image codec was designed for machine consumption and demonstrates significant improvements in object detection and instance segmentation tasks over traditional human-targeted codecs~\cite{le2021image}, which leverages neural networks and customized training strategies to enhance task-specific performance while maintaining efficiency.
\textcolor{black}{
For video understanding under low-bitrate conditions, a comprehensive study has been conducted to fully investigate the promising principles for making compact representation~\cite{tian2024coding}.
}
Recently, researchers have proposed the Collaborative Intelligence (CI)~\cite{bajic2021collaborative,alvar2021pareto} scheme to systematically formulate the compact representation of intermediate features between edge devices and the cloud.
\textcolor{black}{
Notably, to transfer the existing codecs from human perception to machine perception without fine-tuning, a transformer-based framework was proposed to use instance-specific and task-specific prompts, significantly outperforming competing methods~\cite{chen2023transtic}.   
}

The evolution of standards for the compact representation of both traditional handcrafted features and modern deep learning-derived features facilitates enhanced interoperability through a unified bitstream syntax. The MPEG-developed Compact Descriptors for Visual Search (CDVS) standard \cite{duan2015overview} outlines methods for feature extraction and defines the bitstream syntax for still images, incorporating both local \cite{chen2014low, paschalakis2012cdvs, tsai2009location, tsai2012improved} and global \cite{lin2014rate} descriptors derived from handcrafted features. Expanding upon this, the Compact Descriptors for Video Analysis (CDVA) standard \cite{duan2018compact}, also by MPEG, represents the first initiative to standardize visual descriptor compression specifically for video content. CDVA integrates handcrafted and deep learning features, including the novel HNIP approach for condensing CNN features \cite{lin2017hnip}. Reflecting the rapid advancements in deep learning, which have significantly broadened the application spectrum of deep learning features in computer vision, MPEG launched the Video Coding for Machines (VCM) \cite{duan2020video} in 2019 to standardize feature compression for machine vision. 
\textcolor{black}{
Recently, to systematically demonstrate the rapid development of VCM, Yang \textit{et al.}~\cite{yang2024video} conducted a comprehensive survey for compact visual representation compression in intelligent collaborative analytics.
} 
Additionally, the Audio Video Coding Standard group in China is actively pursuing the standardization of compact representations for deep learning features and models, aiming to support the diverse requirements of modern multimedia applications. 

\subsection{Unified Representation for Dynamic Tasks}
\textcolor{black}{
Unified representation frameworks aim to generate a single compressed representation suitable for multiple dynamic tasks. By enabling adaptable data streams that adjust to task-specific requirements, these frameworks eliminate redundant data processing and transmission. In green multimedia, this approach enhances energy efficiency by dynamically aligning resource usage with task needs, reducing the overall data load and processing requirements.
This is aligned with the mechanism of the human brain, where the input signals across different regions and hierarchical levels can be adaptively integrated according to the demands of various tasks.
}
However, the traditional coding targets at the reconstruction of only feature or texture, which cannot support multiple tasks simultaneously. Regarding the diverse goals of downstream tasks, recent efforts have been devoted to scalable layered coding, including features and textures that uniquely contribute to the efficient representation of visual information from different aspects. 
For example, in \cite{wang2019scalable}, the base layer accounts for the compact feature and the enhancement layer reconstructs the texture. 
\textcolor{black}{
Lin \textit{et al.} proposed to divide the bitstream into semantic, structure, and texture layers to achieve both machine and human vision~\cite{lin2023deepsvc}. 
}
Although these integrated approaches are still far from the actual mechanism of HVS, which has not been well understood, they still offer several unique benefits.
For example, the compact representation of high-quality features extracted directly from the original texture could largely enhance the analytical capabilities. 
\textcolor{black}{
Furthermore, at the decoding end, features can be processed directly from the bitstream syntax elements~\cite{zhang2016real}, such as motion vectors or coding modes, which are part of the video coding standard. This strategy significantly reduces the computational load compared to standard video decoding and separate feature extraction, potentially enhancing the performance of downstream tasks such as object detection.
}
Thirdly, the synergy between feature and texture coding improves the overall efficiency of the joint bitstream, since the redundancy among the two representations has been already removed.

The effectiveness of the unified coding approach has been explored in terms of both analytical and reconstructive capabilities, paving the way for advanced unified schemes for visual information compression \cite{ma2018joint}. Zhang \emph{et al.} have shown that encoding SIFT features alongside texture not only preserves but can enhance performance in mobile video applications compared to coding textures alone \cite{zhang2016joint}. Moreover, Li \emph{et al.} introduced the Texture-Feature-Quality-Index (TFQI), which optimizes bit allocation for maximum utility, encompassing both analysis and monitoring purposes \cite{li2018joint}. The concept of Joint Features and Textures Coding (JFTC) has been developed to maximize both visual analysis and compression effectiveness \cite{ma2018joint}. Additionally, Wang \emph{et al.} have examined the interactions between deep learning features and textures within a scalable framework, where the base layer is used for analytics and the enhancement layer augments the unrecoverable texture details. This approach significantly improves both the comprehension and reconstruction of visual signals \cite{wang2019scalable}. Mao \emph{et al.} exploited the priors from pre-trained generative models and proposed a scalable face compression framework, which achieves both promising machine analysis and human perception performance under ultra-low bitrate conditions. 
Sun \emph{et al.} proposed a semantic structure bitstream construction method to code visual data and achieve diverse intelligent tasks at the decoder side~\cite{sun2020semantic}.
By dividing the bitsteam into a base layer and enhanced layer, Wang \emph{et al.} proposed a joint feature and texture representation framework, facilitating multitask learning~\cite{wang2021towards}.

Dynamic green operations not only promote significant energy savings but also enhance the adaptability and efficiency of multimedia systems. By tailoring computational efforts to task needs, these operations optimize resource allocation, minimizing unnecessary power consumption and extending the operational lifespan of devices. 
Furthermore, the dynamic adaptation of multimedia processing mirrors the flexibility and efficiency of biological systems, such as the human brain, which optimizes its functions based on environmental demands and internal states~\cite{padamsey2022neocortex}. By integrating such biologically-inspired approaches, dynamic green operations pave the way for more sustainable and intelligent multimedia technologies.
Specifically, to further elaborate on the potential of scalable coding schemes, Choi \emph{et al.} presented an end-to-end learned image codec optimized for both human viewing and machine vision tasks, which utilizes a layered approach where the base layer is dedicated to simple tasks and additional layers enhance capability for more complex tasks~\cite{choi2022scalable}. 
To jointly leverage the priors from the external database and face sketch, an ultra-low bitrate face compression was developed to reconstruct perceptually meaningful results while providing analytical performance~\cite{mao2023peering}. 
\textcolor{black}{
Notably, Yi \emph{et al.} proposes a task-driven video compression framework optimized for both human and machine vision by incorporating multi-scale motion estimation, multi-frame feature fusion, and reference-based in-loop filters, achieving a balance between signal and semantic fidelity as well as superior rate-distortion and rate-accuracy performance~\cite{10004012}.
}
Recently, Shen \emph{et al.} proposed a versatile neural video coding (VNVC) framework, which not only designs highly compact representations for intermediate features but also enables their generalization in diverse downstream tasks, including reconstruction, enhancement, and analysis~\cite{sheng2024vnvc}. 
Ge \emph{et al.} designed an innovative encoder controller for DVC that employs mode prediction and Group of Pictures (GoP) selection to adapt the pre-trained DVC decoder for various tasks~\cite{ge2024task}.
Li \emph{et al.} introduced a human-machine collaborative image coding framework utilizing Implicit Neural Representations (INR)~\cite{li2024human}. This framework balances high-efficiency visual data compression for human vision with reduced information transmission for machine vision tasks.
Zhang \emph{et al.} proposed a Unified and Scalable Deep Image Compression (USDIC) framework that adopts a novel approach by introducing an Information Splitting Mechanism (ISM) that effectively segregates semantic and visual features for optimized machine analysis and human viewing, respectively~\cite{zhang2024unified}. 

\subsection{Connections with AI and Networking}
\textcolor{black}{
The principles, techniques, and applications of compact visual representation in green multimedia are deeply connected with a number of sectors in networking and AI. In particular, AI-based approaches can enable intelligent decision-making for data prioritization, compression, and transmission, while network optimizations may ensure data is handled in an energy-efficient manner. 
There are four aspects this subsection wants to discuss for the greener multimedia infrastructure connected with AI. 
}

\begin{itemize}
\item \textcolor{black}{Artificial Intelligence Generated Content (AIGC).}
\textcolor{black}{AIGC is expected to yield vast amounts of data with its rapid development in recent years~\cite{cao2023comprehensive, foo2023aigc, xu2024unleashing}. Consequently, a significant challenge is how to represent such data in a compact form, leading to sustainable multimedia. Ideally, the end-users of AIGC could range from human beings to various machine systems.} In this context, the compression of textures and semantic features, along with their effective integration, emerges as a promising area of development. However, AIGC data possesses an inherent machine-born attribute, which allows for the seamless integration of AIGC models into the decoding process, thereby facilitating an extremely compact representation. This integration could be pivotal in enhancing efficiency. However, how to reduce the environmental footprint of data generation and consumption still needs to be further investigated. 

\item \textcolor{black}{Large vision language models (LVLMs)}. Despite LVLMs' significant computational requirements, play a pivotal role in the future of green multimedia. They have the potential to be the ultimate consumers of visual data, necessitating efficient methods for compact representation of such data. The exploration of these methods is of great importance. Research has demonstrated that leveraging LVLMs for optimizing the compression can yield substantial coding gains~\cite{li2024high}. However, the costs associated with their use, particularly on the decoder side, are a significant consideration. However, evidence has shown that the integration of LVLMs into the decoding process could be a key strategy for achieving ultra-compact data representation~\cite{zhang2024video}. In the future, how the light weigh LVLMs can be incorporated into the coding process still needs further exploration. 

\item \textcolor{black}{Knowledge Centric Networking (KCN).}
KCN~\cite{wu2019vision} aims to address the pressing demand for knowledge extraction and dissemination over networks. 
Feature extraction can be treated as the process of distilling descriptive knowledge from raw data collected by acquisition devices or edge sensors. This approach reduces the volume of data that needs to be transmitted, thereby increasing system intelligence and enhancing communication capability at a reduced cost. KCN achieves this by harnessing the network's computing, storage, and communication resources to efficiently produce and disseminate knowledge throughout the network.

\item \textcolor{black}{Edge Computing (EC).}
In principle, the representation that involves data analytics is intrinsically correlated with edge computing, which aims to push the data processing and analytics to edge~\cite{shi2016edge}. In this manner, low response time, latency, and bandwidth consumption can be ensured. The capability of edge computing also benefits the relevant compact representation technologies. For example, in compact feature representation, the feature extraction in the edge end can be subsequently transmitted to the cloud side for information fusion. Moreover, the models at the edge end can be frequently updated with the model communication technologies. 

\end{itemize}
\section{Envisions}

\textcolor{black}{Although green computing for videos has achieved significant progress, there is still a considerable gap compared to the performance of the human visual system. At the same time, the purpose of video representation is to reconstruct rather than to understand visual information as the HVS does. These two different objectives ultimately make it impossible to bridge the performance gap, even if the compression performance is significantly improved in each generation of the video coding standard. 
However, in the future, we believe that new technologies will continue to emerge to promote the development of this field. In particular, the deep learning algorithms, which numerous state-of-the-art compression algorithms rely on, are rather more complicated than the human brain (12-25 watts). Recently, there have also been alternative solutions that attempt to avoid the huge amount of training resources.  
For instance, the subspace approximation with augmented kernels (SAAK) transform, as discussed in~\cite{kuo2018data}, has proven efficient in various processing and pattern recognition tasks. This approach presents a promising avenue for reducing the energy footprint of video processing technologies. 
Nonetheless, attaining high compression efficiency while simultaneously minimizing energy consumption presents challenges. Reduced energy usage typically correlates with fewer processing operations, leading to diminished video quality. Moreover, existing hardware may not be optimized for energy-efficient processing, necessitating innovations in both software and hardware design. These challenges may serve as key motivators for pioneering research in the realm of green computing for videos.}

\textcolor{black}{Another promising direction is communication among AI agents, which will probably open up a fascinating frontier in the field of artificial intelligence and information communication. Semantic communication~\cite{luo2022semantic}, which is a long-standing research topic, has been widely studied in the literature recently, paving the way toward the next generation of communication. However, the process of achieving brain-like semantic communication for AI agents is still in its infancy. This form of communication focuses on how to create the language for machines, and translators between different machine languages. 
The linguistic structure among AI agents requires a redesign, a task that is likely best suited for AI capabilities. 
However, without any doubt, such communication is greener than human-centric communication, and AI agent systems are expected to interact seamlessly within this. 
Furthermore, ensuring that communications between AI agents are not only efficient but also secure and interpretable by human overseers to prevent miscommunications and potential misuse is also a challenge for developing a machine language that can encapsulate complex ideas succinctly and accurately.}

\textcolor{black}{
Moreover, neuromorphic computing~\cite{markovic2020physics,shastri2021photonics} is also expected to greatly enhance communication efficiency, leading to green communication. Neuromorphic systems which are typically designed to mimic the neural structure and functionalities of brains, could potentially allow for more efficient communication between AI agents, as well as between humans and AI agents.
However, building systems that accurately mimic human neural architectures is technically challenging and demands innovations in materials science and electronic engineering. Additionally, the high cost of developing and producing neuromorphic chips may impede their broad adoption and commercial integration, which necessitates researchers to explore relevant fields. Besides, one potential research direction of neuromorphic computing is to integrate it with existing AI frameworks to enhance their computational efficiency and reduce power consumption. 
Furthermore, developing scalable neuromorphic systems that can be scaled up efficiently to handle complex computations akin to those required for real-time AI communications, can significantly contribute to more sustainable and efficient computing and communication solutions.
} 

\section{Conclusions}
In this survey, we have explored the progression and future directions of compact visual data representation within the context of green multimedia systems. We have demonstrated how insights from the HVS can be integrated into digital systems to achieve not only higher compression ratios but also enhanced energy efficiency. Our discussion encompassed compact video compression, in particular the perceptual coding and cloud based coding which leverage the perceptual characteristics and memory of HVS. 
In terms of compact feature compression, the push towards front-end intelligence has fostered developments in compact feature representations that support efficient downstream tasks on edge devices. 
Moreover, the concept of unified perception, which combines texture and feature coding, was highlighted as a novel approach to meet the dual needs of human perception and machine-based analytics. This strategy not only preserves the quality of visual information but also simplifies the processing and analysis, paving the way for more efficient multimedia systems.

As AI and machine learning continue to evolve, alongside advancements in edge computing, these technologies promise to further reduce energy consumption and enhance the efficiency of multimedia data handling. The path forward involves bridging the current capabilities of digital systems with the biological efficiency of the HVS, which will be pivotal in ushering in an era of truly green multimedia technologies. The continuing research in this domain is expected to focus on narrowing the performance gap between digital systems and the HVS, thus fostering a more sustainable approach to green multimedia technology development.

\bibliographystyle{IEEEtran}
\bibliography{refs}

\begin{thebibliography}{100}
\providecommand{\url}[1]{#1}
\csname url@samestyle\endcsname
\providecommand{\newblock}{\relax}
\providecommand{\bibinfo}[2]{#2}
\providecommand{\BIBentrySTDinterwordspacing}{\spaceskip=0pt\relax}
\providecommand{\BIBentryALTinterwordstretchfactor}{4}
\providecommand{\BIBentryALTinterwordspacing}{\spaceskip=\fontdimen2\font plus
\BIBentryALTinterwordstretchfactor\fontdimen3\font minus \fontdimen4\font\relax}
\providecommand{\BIBforeignlanguage}[2]{{%
\expandafter\ifx\csname l@#1\endcsname\relax
\typeout{** WARNING: IEEEtran.bst: No hyphenation pattern has been}%
\typeout{** loaded for the language `#1'. Using the pattern for}%
\typeout{** the default language instead.}%
\else
\language=\csname l@#1\endcsname
\fi
#2}}
\providecommand{\BIBdecl}{\relax}
\BIBdecl

\bibitem{chen2020internet}
C.~W. Chen, ``Internet of video things: Next-generation iot with visual sensors,'' \emph{IEEE Internet of Things Journal}, vol.~7, no.~8, pp. 6676--6685, 2020.

\bibitem{foo2023ai}
L.~G. Foo, H.~Rahmani, and J.~Liu, ``{AI}-generated content ({AIGC}) for various data modalities: A survey,'' \emph{arXiv preprint arXiv:2308.14177}, vol.~2, p.~2, 2023.

\bibitem{wiegand2003overview}
T.~Wiegand, G.~J. Sullivan, G.~Bjontegaard, and A.~Luthra, ``Overview of the {H.264/AVC} video coding standard,'' \emph{IEEE Transactions on Circuits and Systems for Video Technology}, vol.~13, no.~7, pp. 560--576, 2003.

\bibitem{sullivan2012overview}
G.~J. Sullivan, J.-R. Ohm, W.-J. Han, and T.~Wiegand, ``Overview of the high efficiency video coding {(HEVC)} standard,'' \emph{IEEE Transactions on Circuits and Systems for Video Technology}, vol.~22, no.~12, pp. 1649--1668, 2012.

\bibitem{bross2021overview}
B.~Bross, Y.-K. Wang, Y.~Ye, S.~Liu, J.~Chen, G.~J. Sullivan, and J.-R. Ohm, ``Overview of the versatile video coding {(VVC)} standard and its applications,'' \emph{IEEE Transactions on Circuits and Systems for Video Technology}, vol.~31, no.~10, pp. 3736--3764, 2021.

\bibitem{chen2018overview}
Y.~Chen, D.~Murherjee, J.~Han, A.~Grange, Y.~Xu, Z.~Liu, S.~Parker, C.~Chen, H.~Su, U.~Joshi \emph{et~al.}, ``An overview of core coding tools in the av1 video codec,'' in \emph{2018 Picture Coding Symposium (PCS)}.\hskip 1em plus 0.5em minus 0.4em\relax IEEE, 2018, pp. 41--45.

\bibitem{8954503}
J.~{Zhang}, C.~{Jia}, M.~{Lei}, S.~{Wang}, S.~{Ma}, and W.~{Gao}, ``Recent development of {AVS} video coding standard: {AVS3},'' in \emph{2019 Picture Coding Symposium (PCS)}, 2019, pp. 1--5.

\bibitem{shannon1948mathematical}
C.~E. Shannon, ``A mathematical theory of communication,'' \emph{The Bell System Technical Journal}, vol.~27, no.~3, pp. 379--423, 1948.

\bibitem{gao2021digital}
W.~Gao, S.~Ma, L.~Duan, Y.~Tian, P.~Xing, Y.~Wang, S.~Wang, H.~Jia, and T.~Huang, ``Digital retina: A way to make the city brain more efficient by visual coding,'' \emph{IEEE Transactions on Circuits and Systems for Video Technology}, 2021.

\bibitem{zhang2023rethinking}
P.~Zhang, S.~Wang, M.~Wang, J.~Li, X.~Wang, and S.~Kwong, ``Rethinking semantic image compression: Scalable representation with cross-modality transfer,'' \emph{IEEE Transactions on Circuits and Systems for Video Technology}, vol.~33, no.~8, pp. 4441--4445, 2023.

\bibitem{herglotz2023extended}
C.~Herglotz, M.~Kränzler, X.~Chu, E.~Fran{\c{c}}ois, Y.~He, and A.~Kaup, ``Extended signaling methods for reduced video decoder power consumption using green metadata,'' \emph{IEEE Transactions on Circuits and Systems II: Express Briefs}, vol.~71, no.~3, pp. 1141--1145, 2024.

\bibitem{herglotz2024energy}
C.~Herglotz, S.~Le~Moan, and A.~Mercat, ``Energy reduction opportunities in {HDR} video encoding,'' in \emph{2024 IEEE International Conference on Image Processing (ICIP)}.\hskip 1em plus 0.5em minus 0.4em\relax IEEE, 2024, pp. 3654--3660.

\bibitem{herglotz2024complexity}
C.~Herglotz, M.~Kr{\"a}nzler, R.~Dai, and A.~Kaup, ``Complexity metrics for {VVC} decoder power reduction in green metadata,'' in \emph{2024 Picture Coding Symposium (PCS)}.\hskip 1em plus 0.5em minus 0.4em\relax IEEE, 2024, pp. 1--5.

\bibitem{kruger2012deep}
N.~Kruger, P.~Janssen, S.~Kalkan, M.~Lappe, A.~Leonardis, J.~Piater, A.~J. Rodriguez-Sanchez, and L.~Wiskott, ``Deep hierarchies in the primate visual cortex: What can we learn for computer vision?'' \emph{IEEE Transactions on Pattern Analysis and Machine Intelligence}, vol.~35, no.~8, pp. 1847--1871, 2013.

\bibitem{brain_report}
\BIBentryALTinterwordspacing
S.~R. Corporation, ``Decadal plan for semiconductors.'' 2021. [Online]. Available: \url{https://www.src.org/about/decadal-plan/decadal-plan-full-report.pdf}
\BIBentrySTDinterwordspacing

\bibitem{zhang2020system}
Y.~Zhang, P.~Qu, Y.~Ji, W.~Zhang, G.~Gao, G.~Wang, S.~Song, G.~Li, W.~Chen, W.~Zheng \emph{et~al.}, ``A system hierarchy for brain-inspired computing,'' \emph{Nature}, vol. 586, no. 7829, pp. 378--384, 2020.

\bibitem{duan2020video}
L.~Duan, J.~Liu, W.~Yang, T.~Huang, and W.~Gao, ``Video coding for machines: A paradigm of collaborative compression and intelligent analytics,'' \emph{IEEE Transactions on Image Processing}, vol.~29, pp. 8680--8695, 2020.

\bibitem{chen2024generative}
B.~Chen, S.~Yin, P.~Chen, S.~Wang, and Y.~Ye, ``Generative visual compression: A review,'' in \emph{2024 IEEE International Conference on Image Processing (ICIP)}, 2024, pp. 3709--3715.

\bibitem{wang2006modern}
Z.~Wang and A.~C. Bovik, ``Modern image quality assessment,'' Ph.D. dissertation, Springer, 2006.

\bibitem{remington2011clinical}
L.~A. Remington and D.~Goodwin, \emph{Clinical anatomy of the visual system E-Book}.\hskip 1em plus 0.5em minus 0.4em\relax Elsevier Health Sciences, 2011.

\bibitem{sziklai1956some}
G.~Sziklai, ``Some studies in the speed of visual perception,'' \emph{IRE Transactions on Information Theory}, vol.~2, no.~3, pp. 125--128, 1956.

\bibitem{lou2019towards}
Y.~Lou, L.-Y. Duan, Y.~Luo, Z.~Chen, T.~Liu, S.~Wang, and W.~Gao, ``Towards digital retina in smart cities: A model generation, utilization and communication paradigm,'' in \emph{2019 IEEE International Conference on Multimedia and Expo (ICME)}.\hskip 1em plus 0.5em minus 0.4em\relax IEEE, 2019, pp. 19--24.

\bibitem{lou2020towards}
------, ``Towards efficient front-end visual sensing for digital retina: A model-centric paradigm,'' \emph{IEEE Transactions on Multimedia}, vol.~22, no.~11, pp. 3002--3013, 2020.

\bibitem{eichermuller2024svt}
L.~Eicherm{\"u}ller, G.~Chaudhari, I.~Katsavounidis, Z.~Lei, H.~Tmar, C.~Herglotz, and A.~Kaup, ``{SVT-AV1} encoding bitrate estimation using motion search information,'' in \emph{2024 32nd European Signal Processing Conference (EUSIPCO)}.\hskip 1em plus 0.5em minus 0.4em\relax IEEE, 2024, pp. 937--941.

\bibitem{8712764}
M.~Wang, J.~Li, L.~Zhang, K.~Zhang, H.~Liu, S.~Wang, S.~Kwong, and S.~Ma, ``Extended quad-tree partitioning for future video coding,'' in \emph{2019 Data Compression Conference (DCC)}, 2019, pp. 300--309.

\bibitem{8917942}
------, ``Extended coding unit partitioning for future video coding,'' \emph{IEEE Transactions on Image Processing}, vol.~29, pp. 2931--2946, 2020.

\bibitem{9115218}
J.~Li, M.~Wang, L.~Zhang, K.~Zhang, H.~Liu, S.~Wang, S.~Ma, and W.~Gao, ``Unified intra mode coding based on short and long range correlations,'' \emph{IEEE Transactions on Image Processing}, vol.~29, pp. 7245--7260, 2020.

\bibitem{Modes67}
N.~Choi, Y.~Piao, K.~Choi, and C.~Kim, ``{CE} 3.3 related: Intra 67 modes coding with 3 {MPM},'' \emph{Joint Video Exploration Team (JVET), doc. JVET-K0529}, Jul. 2018.

\bibitem{zhang2018enhanced}
K.~Zhang, J.~Chen, L.~Zhang, X.~Li, and M.~Karczewicz, ``Enhanced cross-component linear model for chroma intra-prediction in video coding,'' \emph{IEEE Transactions on Image Processing}, vol.~27, no.~8, pp. 3983--3997, 2018.

\bibitem{9105802}
J.~{Li}, M.~{Wang}, L.~{Zhang}, K.~{Zhang}, S.~{Wang}, S.~{Wang}, S.~{Ma}, and W.~{Gao}, ``Sub-sampled cross-component prediction for chroma component coding,'' in \emph{2020 Data Compression Conference (DCC)}, 2020, pp. 203--212.

\bibitem{9105965}
J.~{Li}, L.~{Zhang}, K.~{Zhang}, H.~{Liu}, M.~{Wang}, S.~{Wang}, S.~{Ma}, and W.~{Gao}, ``Prediction with multi-cross component,'' in \emph{2020 IEEE International Conference on Multimedia Expo Workshops (ICMEW)}, 2020, pp. 1--6.

\bibitem{IBC_AVS3}
Y.~{Wang}, X.~{Xu}, and S.~{Liu}, ``Intra block copy in {AVS3} video coding standard,'' in \emph{2020 IEEE International Conference on Multimedia Expo Workshops (ICMEW)}, 2020, pp. 1--6.

\bibitem{HBVP}
W.~{Yin}, J.~{Xu}, L.~{Zhang}, K.~{Zhang}, H.~{Liu}, and X.~{Fan}, ``History based block vector predictor for intra block copy,'' in \emph{2020 IEEE International Conference on Multimedia Expo Workshops (ICMEW)}, 2020, pp. 1--6.

\bibitem{PLT}
Y.~{Sun}, J.~{Lou}, Y.~{Chao}, H.~{Wang}, V.~{Seregin}, and M.~{Karczewicz}, ``Analysis of palette mode on versatile video coding,'' in \emph{2019 IEEE Conference on Multimedia Information Processing and Retrieval (MIPR)}, 2019, pp. 455--458.

\bibitem{StringCopy}
Q.~Zhou, L.~Zhao, K.~Zhou, T.~Lin, H.~Wang, S.~Wang, and M.~Jiao, ``String prediction for 4:2:0 format screen content coding and its implementation in {AVS3},'' \emph{IEEE Transactions on Multimedia}, vol.~23, pp. 3867--3876, 2021.

\bibitem{affine_avs}
T.~{Fu}, K.~{Zhang}, H.~{Liu}, L.~{Zhang}, S.~{Wang}, S.~{Ma}, and W.~{Gao}, ``Affine direct/skip mode with motion vector differences in video coding,'' in \emph{2020 IEEE International Conference on Multimedia Expo Workshops (ICMEW)}, 2020, pp. 1--6.

\bibitem{affine_tip}
K.~{Zhang}, Y.~{Chen}, L.~{Zhang}, W.~{Chien}, and M.~{Karczewicz}, ``An improved framework of affine motion compensation in video coding,'' \emph{IEEE Transactions on Image Processing}, vol.~28, no.~3, pp. 1456--1469, 2019.

\bibitem{amvr}
H.~{Liu}, L.~{Zhang}, K.~{Zhang}, J.~{Xu}, Y.~{Wang}, J.~{Luo}, and Y.~{He}, ``Adaptive motion vector resolution for affine-inter mode coding,'' in \emph{2019 Picture Coding Symposium (PCS)}, 2019, pp. 1--4.

\bibitem{HMVP_DCC}
L.~{Zhang}, K.~{Zhang}, H.~{Liu}, H.~C. {Chuang}, Y.~{Wang}, J.~{Xu}, P.~{Zhao}, and D.~{Hong}, ``History-based motion vector prediction in versatile video coding,'' in \emph{2019 Data Compression Conference (DCC)}, 2019, pp. 43--52.

\bibitem{mccann2010samsung}
K.~McCann, W.-J. Han, I.-K. Kim, J.-H. Min, E.~Alshina, A.~Alshin, T.~Lee, J.~Chen, V.~Seregin, S.~Lee \emph{et~al.}, ``Samsung's response to the call for proposals on video compression technology,'' \emph{JCTVC-A124}, pp. 1--42, 2010.

\bibitem{alshin2010bi}
A.~Alshin, E.~Alshina, and T.~Lee, ``Bi-directional optical flow for improving motion compensation,'' in \emph{28th Picture Coding Symposium}.\hskip 1em plus 0.5em minus 0.4em\relax IEEE, 2010, pp. 422--425.

\bibitem{jeong2020merge}
S.~Jeong, Y.~Piao, M.~W. Park, M.~Park, A.~Tamse, N.~Choi, K.~Choi, W.~Choi, and C.~Kim, ``Merge mode with motion vector difference,'' in \emph{2020 IEEE International Conference on Image Processing (ICIP)}.\hskip 1em plus 0.5em minus 0.4em\relax IEEE, 2020, pp. 1157--1160.

\bibitem{DMVR_CSVT}
H.~Gao, X.~Chen, S.~Esenlik, J.~Chen, and E.~Steinbach, ``Decoder-side motion vector refinement in {VVC}: Algorithm and hardware implementation considerations,'' \emph{IEEE Transactions on Circuits and Systems for Video Technology}, vol.~31, no.~8, pp. 3197--3211, 2021.

\bibitem{Trellis_DCC}
H.~Schwarz, T.~Nguyen, D.~Marpe, and T.~Wiegand, ``Hybrid video coding with trellis-coded quantization,'' in \emph{2019 Data Compression Conference (DCC)}, March 2019, pp. 182--191.

\bibitem{TransformVVC}
X.~Zhao, J.~Chen, M.~Karczewicz, A.~Said, and V.~Seregin, ``Joint separable and non-separable transforms for next-generation video coding,'' \emph{IEEE Transactions on Image Processing}, vol.~27, no.~5, pp. 2514--2525, May 2018.

\bibitem{IST_ICMEW}
Y.~{Zhang}, K.~{Zhang}, L.~{Zhang}, H.~{Liu}, Y.~{Wang}, S.~{Wang}, S.~{Ma}, and W.~{Gao}, ``Implicit-selected transform in video coding,'' in \emph{2020 IEEE International Conference on Multimedia Expo Workshops (ICMEW)}, 2020, pp. 1--6.

\bibitem{orchard1994overlapped}
M.~T. Orchard and G.~J. Sullivan, ``Overlapped block motion compensation: An estimation-theoretic approach,'' \emph{IEEE Transactions on Image Processing}, vol.~3, no.~5, pp. 693--699, 1994.

\bibitem{HEVC_CABAC}
V.~Sze and M.~Budagavi, ``{High Throughput CABAC Entropy Coding in HEVC},'' \emph{IEEE Transactions on Circuits and Systems for Video Technology}, vol.~22, no.~12, pp. 1778--1791, Dec 2012.

\bibitem{8713491}
B.~Xiao, H.~Wang, J.~Wu, S.~Kwong, and C.-C.~J. Kuo, ``A multi-grained parallel solution for hevc encoding on heterogeneous platforms,'' \emph{IEEE Transactions on Multimedia}, vol.~21, no.~12, pp. 2997--3009, 2019.

\bibitem{youvalari2023filtered}
R.~G. Youvalari, D.~B. Sansli, P.~Astola, and J.~Lainema, ``Filtered intra template matching prediction for future video coding,'' in \emph{2023 31st European Signal Processing Conference (EUSIPCO)}.\hskip 1em plus 0.5em minus 0.4em\relax IEEE, 2023, pp. 576--579.

\bibitem{abdoli2024video}
M.~Abdoli, R.~G. Youvalari, K.~Naser, K.~Reuz{\'e}, and F.~L. L{\'e}annec, ``Video compression beyond vvc: Quantitative analysis of intra coding tools in enhanced compression model {(ECM)},'' \emph{arXiv preprint arXiv:2404.07872}, 2024.

\bibitem{pu2024intra}
F.~Pu, T.~Lu, P.~Yin, S.~McCarthy, J.~R. Arumugam, A.~Natesan, V.~Valvaiker, J.~N. Shingala, X.~Li, R.-L. Liao \emph{et~al.}, ``Intra template matching prediction with fusion techniques,'' in \emph{2024 Data Compression Conference (DCC)}.\hskip 1em plus 0.5em minus 0.4em\relax IEEE, 2024, pp. 93--102.

\bibitem{liao2024improvement}
R.-L. Liao, J.~Chen, Y.~Ye, and X.~Li, ``An improvement to subblock-based temporal motion vector prediction beyond {VVC},'' in \emph{2024 Data Compression Conference (DCC)}.\hskip 1em plus 0.5em minus 0.4em\relax IEEE, 2024, pp. 83--92.

\bibitem{deng2024inter}
Z.~Deng, K.~Zhang, and L.~Zhang, ``Inter cross-component prediction merge mode for video coding beyond {VVC},'' in \emph{2024 Data Compression Conference (DCC)}.\hskip 1em plus 0.5em minus 0.4em\relax IEEE, 2024, pp. 551--551.

\bibitem{vishwanath2023discrete}
B.~Vishwanath, K.~Zhang, and L.~Zhang, ``A discrete-mapping-based cross-component prediction paradigm for screen content coding,'' \emph{IEEE Transactions on Image Processing}, vol.~33, pp. 16--26, 2023.

\bibitem{berger2017survey}
M.~{Berger et al}, ``A survey of surface reconstruction from point clouds,'' in \emph{Computer Graphics Forum}, vol.~36, no.~1, 2017, pp. 301--329.

\bibitem{schwarz2018emerging}
S.~Schwarz, M.~Preda, V.~Baroncini, M.~Budagavi, P.~Cesar, P.~A. Chou, R.~A. Cohen, M.~Krivoku{\'c}a, S.~Lasserre, Z.~Li \emph{et~al.}, ``Emerging mpeg standards for point cloud compression,'' \emph{IEEE Journal on Emerging and Selected Topics in Circuits and Systems}, vol.~9, no.~1, pp. 133--148, 2018.

\bibitem{maglo20153d}
A.~Maglo, G.~Lavou{\'e}, F.~Dupont, and C.~Hudelot, ``3d mesh compression: Survey, comparisons, and emerging trends,'' \emph{ACM Computing Surveys (CSUR)}, vol.~47, no.~3, pp. 1--41, 2015.

\bibitem{zhou2019video}
Y.~Zhou, L.~Tian, C.~Zhu, X.~Jin, and Y.~Sun, ``Video coding optimization for virtual reality 360-degree source,'' \emph{IEEE Journal of Selected Topics in Signal Processing}, vol.~14, no.~1, pp. 118--129, 2019.

\bibitem{mildenhall2021nerf}
B.~Mildenhall, P.~P. Srinivasan, M.~Tancik, J.~T. Barron, R.~Ramamoorthi, and R.~Ng, ``Nerf: Representing scenes as neural radiance fields for view synthesis,'' \emph{Communications of the ACM}, vol.~65, no.~1, pp. 99--106, 2021.

\bibitem{kerbl20233d}
B.~Kerbl, G.~Kopanas, T.~Leimk{\"u}hler, and G.~Drettakis, ``3d gaussian splatting for real-time radiance field rendering.'' \emph{ACM Transactions on Graphics}, vol.~42, no.~4, pp. 139--1, 2023.

\bibitem{jang2019video}
E.~S. Jang, M.~Preda, K.~Mammou, A.~M. Tourapis, J.~Kim, D.~B. Graziosi, S.~Rhyu, and M.~Budagavi, ``Video-based point-cloud-compression standard in {MPEG}: From evidence collection to committee draft [standards in a nutshell],'' \emph{IEEE Signal Processing Magazine}, vol.~36, no.~3, pp. 118--123, 2019.

\bibitem{li2020efficient}
L.~Li, Z.~Li, S.~Liu, and H.~Li, ``Efficient projected frame padding for video-based point cloud compression,'' \emph{IEEE Transactions on Multimedia}, vol.~23, pp. 2806--2819, 2020.

\bibitem{li2019advanced}
L.~Li, Z.~Li, V.~Zakharchenko, J.~Chen, and H.~Li, ``Advanced {3D} motion prediction for video-based dynamic point cloud compression,'' \emph{IEEE Trans. Image Process.}, vol.~29, pp. 289--302, 2019.

\bibitem{akhtar2021video}
A.~Akhtar, W.~Gao, L.~Li, Z.~Li, W.~Jia, and S.~Liu, ``Video-based point cloud compression artifact removal,'' \emph{IEEE Transactions on Multimedia}, vol.~24, pp. 2866--2876, 2022.

\bibitem{graziosi2020overview}
D.~Graziosi, O.~Nakagami, S.~Kuma, A.~Zaghetto, T.~Suzuki, and A.~Tabatabai, ``An overview of ongoing point cloud compression standardization activities: Video-based {(V-PCC)} and geometry-based {(G-PCC)},'' \emph{APSIPA Transactions on Signal and Information Processing}, vol.~9, 2020.

\bibitem{ainala2016improved}
K.~{Ainala et al}, ``An improved enhancement layer for octree based point cloud compression with plane projection approximation,'' in \emph{Appl. digit. image process. XXXIX}, vol. 9971.\hskip 1em plus 0.5em minus 0.4em\relax SPIE, 2016, pp. 223--231.

\bibitem{dziembowski2023immersive}
A.~Dziembowski, D.~Mieloch, J.~Y. Jeong, and G.~Lee, ``Immersive video postprocessing for efficient video coding,'' \emph{IEEE Transactions on Circuits and Systems for Video Technology}, vol.~33, no.~8, pp. 4349--4361, 2023.

\bibitem{balle2016end}
J.~Ball{\'e}, V.~Laparra, and E.~Simoncelli, ``End-to-end optimized image compression,'' in \emph{International Conference on Learning Representations}, 2016.

\bibitem{balle2016end2}
J.~Ball{\'e}, V.~Laparra, and E.~P. Simoncelli, ``End-to-end optimization of nonlinear transform codes for perceptual quality,'' in \emph{2016 Picture Coding Symposium (PCS)}.\hskip 1em plus 0.5em minus 0.4em\relax IEEE, 2016, pp. 1--5.

\bibitem{balle2018variational}
J.~Ball{\'e}, D.~Minnen, S.~Singh, S.~J. Hwang, and N.~Johnston, ``Variational image compression with a scale hyperprior,'' in \emph{International Conference on Learning Representations}, 2018.

\bibitem{cheng2020learned}
Z.~Cheng, H.~Sun, M.~Takeuchi, and J.~Katto, ``Learned image compression with discretized gaussian mixture likelihoods and attention modules,'' in \emph{Proceedings of the IEEE/CVF Conference on Computer Vision and Pattern Recognition}, 2020, pp. 7939--7948.

\bibitem{chang2020conceptual}
J.~Chang, Z.~Zhao, C.~Jia, S.~Wang, L.~Yang, Q.~Mao, J.~Zhang, and S.~Ma, ``Conceptual compression via deep structure and texture synthesis,'' \emph{IEEE Transactions on Image Processing}, vol.~31, pp. 2809--2823, 2022.

\bibitem{9858899}
Z.~Tang, H.~Wang, X.~Yi, Y.~Zhang, S.~Kwong, and C.-C.~J. Kuo, ``Joint graph attention and asymmetric convolutional neural network for deep image compression,'' \emph{IEEE Transactions on Circuits and Systems for Video Technology}, vol.~33, no.~1, pp. 421--433, 2023.

\bibitem{zhao2024neural}
Y.~Zhao, W.~He, C.~Jia, Q.~Wang, J.~Li, Y.~Li, C.~Lin, K.~Zhang, L.~Zhang, and S.~Ma, ``A neural-network enhanced video coding framework beyond {ECM},'' in \emph{2024 Data Compression Conference (DCC)}, 2024, pp. 605--605.

\bibitem{lu2019dvc}
G.~Lu, W.~Ouyang, D.~Xu, X.~Zhang, C.~Cai, and Z.~Gao, ``{DVC}: An end-to-end deep video compression framework,'' in \emph{2019 IEEE/CVF Conference on Computer Vision and Pattern Recognition (CVPR)}, 2019, pp. 10\,998--11\,007.

\bibitem{lu2020end}
G.~Lu, X.~Zhang, W.~Ouyang, L.~Chen, Z.~Gao, and D.~Xu, ``An end-to-end learning framework for video compression,'' \emph{IEEE Transactions on Pattern Analysis and Machine Intelligence}, vol.~43, no.~10, pp. 3292--3308, 2021.

\bibitem{yang2020learning}
R.~Yang, F.~Mentzer, L.~V. Gool, and R.~Timofte, ``Learning for video compression with hierarchical quality and recurrent enhancement,'' in \emph{Proceedings of the IEEE/CVF Conference on Computer Vision and Pattern Recognition}, 2020, pp. 6628--6637.

\bibitem{li2021deep}
J.~Li, B.~Li, and Y.~Lu, ``Deep contextual video compression,'' \emph{Advances in Neural Information Processing Systems}, vol.~34, pp. 18\,114--18\,125, 2021.

\bibitem{sheng2022temporal}
X.~Sheng, J.~Li, B.~Li, L.~Li, D.~Liu, and Y.~Lu, ``Temporal context mining for learned video compression,'' \emph{IEEE Transactions on Multimedia}, vol.~25, pp. 7311--7322, 2022.

\bibitem{li2022hybrid}
J.~Li, B.~Li, and Y.~Lu, ``Hybrid spatial-temporal entropy modelling for neural video compression,'' in \emph{Proceedings of the 30th ACM International Conference on Multimedia}, 2022, pp. 1503--1511.

\bibitem{li2023neural}
------, ``Neural video compression with diverse contexts,'' in \emph{Proceedings of the IEEE/CVF Conference on Computer Vision and Pattern Recognition}, 2023, pp. 22\,616--22\,626.

\bibitem{li2024neural}
------, ``Neural video compression with feature modulation,'' in \emph{Proceedings of the IEEE/CVF Conference on Computer Vision and Pattern Recognition}, 2024, pp. 26\,099--26\,108.

\bibitem{sheng2024nvc}
X.~Sheng, C.~Tang, L.~Li, D.~Liu, and F.~Wu, ``{NVC-1B}: A large neural video coding model,'' \emph{arXiv preprint arXiv:2407.19402}, 2024.

\bibitem{guo-hua2023evc}
\BIBentryALTinterwordspacing
W.~Guo-Hua, J.~Li, B.~Li, and Y.~Lu, ``{EVC}: Towards real-time neural image compression with mask decay,'' in \emph{The Eleventh International Conference on Learning Representations}, 2023. [Online]. Available: \url{https://openreview.net/forum?id=XUxad2Gj40n}
\BIBentrySTDinterwordspacing

\bibitem{10.1145/3650034}
\BIBentryALTinterwordspacing
R.~Yang, D.~Liu, S.~Ma, F.~Wu, and W.~Gao, ``Perceptual quality-oriented rate allocation via distillation from end-to-end image compression,'' \emph{ACM Trans. Multimedia Comput. Commun. Appl.}, vol.~20, no.~7, Apr. 2024. [Online]. Available: \url{https://doi.org/10.1145/3650034}
\BIBentrySTDinterwordspacing

\bibitem{10.1145/3652148}
\BIBentryALTinterwordspacing
S.~Huo, D.~Liu, H.~Zhang, L.~Li, S.~Ma, F.~Wu, and W.~Gao, ``Towards hybrid-optimization video coding,'' \emph{ACM Comput. Surv.}, vol.~56, no.~9, Apr. 2024. [Online]. Available: \url{https://doi.org/10.1145/3652148}
\BIBentrySTDinterwordspacing

\bibitem{kannur2009power}
A.~K. Kannur and B.~Li, ``Power-aware content-adaptive {H.264} video encoding,'' in \emph{2009 IEEE International Conference on Acoustics, Speech and Signal Processing}.\hskip 1em plus 0.5em minus 0.4em\relax IEEE, 2009, pp. 925--928.

\bibitem{hemmati2013game}
M.~Hemmati, A.~Javadtalab, A.~A. Nazari~Shirehjini, S.~Shirmohammadi, and T.~Arici, ``Game as video: Bit rate reduction through adaptive object encoding,'' in \emph{Proceeding of the 23rd ACM Workshop on Network and Operating Systems Support for Digital Audio and Video}, 2013, pp. 7--12.

\bibitem{mazumdar2019perceptual}
A.~Mazumdar, B.~Haynes, M.~Balazinska, L.~Ceze, A.~Cheung, and M.~Oskin, ``Perceptual compression for video storage and processing systems,'' in \emph{Proceedings of the ACM Symposium on Cloud Computing}, 2019, pp. 179--192.

\bibitem{wu2013perceptual}
H.~R. Wu, A.~R. Reibman, W.~Lin, F.~Pereira, and S.~S. Hemami, ``Perceptual visual signal compression and transmission,'' \emph{Proceedings of the IEEE}, vol. 101, no.~9, pp. 2025--2043, 2013.

\bibitem{zhang2023survey}
Y.~Zhang, L.~Zhu, G.~Jiang, S.~Kwong, and C.-C.~J. Kuo, ``A survey on perceptually optimized video coding,'' \emph{ACM Computing Surveys}, vol.~55, no.~12, pp. 1--37, 2023.

\bibitem{zhang2011fsim}
L.~Zhang, L.~Zhang, X.~Mou, and D.~Zhang, ``{FSIM}: A feature similarity index for image quality assessment,'' \emph{IEEE transactions on Image Processing}, vol.~20, no.~8, pp. 2378--2386, 2011.

\bibitem{zeng2014characterizing}
K.~Zeng, T.~Zhao, A.~Rehman, and Z.~Wang, ``Characterizing perceptual artifacts in compressed video streams,'' in \emph{Human vision and electronic imaging XIX}, vol. 9014.\hskip 1em plus 0.5em minus 0.4em\relax SPIE, 2014, pp. 173--182.

\bibitem{vmaf}
\BIBentryALTinterwordspacing
Z.~Li, A.~Anne, K.~Ioannis, M.~Anush, and M.~Megha, ``Toward a practical perceptual video quality metric,'' 2016. [Online]. Available: \url{https://netflixtechblog.com/toward-a-practical-perceptual-video-quality-metric-653f208b9652}
\BIBentrySTDinterwordspacing

\bibitem{luo2021vmaf}
Z.~Luo, C.~Zhu, Y.~Huang, R.~Xie, L.~Song, and C.-C.~J. Kuo, ``{VMAF} oriented perceptual coding based on piecewise metric coupling,'' \emph{IEEE Transactions on Image Processing}, vol.~30, pp. 5109--5121, 2021.

\bibitem{zhang2018unreasonable}
R.~Zhang, P.~Isola, A.~A. Efros, E.~Shechtman, and O.~Wang, ``The unreasonable effectiveness of deep features as a perceptual metric,'' in \emph{2018 IEEE/CVF Conference on Computer Vision and Pattern Recognition}, 2018, pp. 586--595.

\bibitem{yang2005motion}
X.~Yang, W.~Lin, Z.~Lu, E.~Ong, and S.~Yao, ``Motion-compensated residue preprocessing in video coding based on just-noticeable-distortion profile,'' \emph{IEEE Transactions on Circuits and Systems for Video Technology}, vol.~15, no.~6, pp. 742--752, 2005.

\bibitem{ki2018learning}
S.~Ki, S.-H. Bae, M.~Kim, and H.~Ko, ``Learning-based just-noticeable-quantization-distortion modeling for perceptual video coding,'' \emph{IEEE Transactions on Image Processing}, vol.~27, no.~7, pp. 3178--3193, 2018.

\bibitem{tian2020just}
T.~Tian, H.~Wang, L.~Zuo, C.-C.~J. Kuo, and S.~Kwong, ``Just noticeable difference level prediction for perceptual image compression,'' \emph{IEEE Transactions on Broadcasting}, vol.~66, no.~3, pp. 690--700, 2020.

\bibitem{nami2022bl}
S.~Nami, F.~Pakdaman, M.~R. Hashemi, and S.~Shirmohammadi, ``{BL-JUNIPER: A CNN-assisted} framework for perceptual video coding leveraging block-level jnd,'' \emph{IEEE Transactions on Multimedia}, vol.~25, pp. 5077--5092, 2022.

\bibitem{241504}
N.~Jayant, J.~Johnston, and R.~Safranek, ``Signal compression based on models of human perception,'' \emph{Proceedings of the IEEE}, vol.~81, no.~10, pp. 1385--1422, 1993.

\bibitem{hadizadeh2013saliency}
H.~Hadizadeh and I.~V. Baji{\'c}, ``Saliency-aware video compression,'' \emph{IEEE Transactions on Image Processing}, vol.~23, no.~1, pp. 19--33, 2013.

\bibitem{wang2011ssim}
S.~Wang, A.~Rehman, Z.~Wang, S.~Ma, and W.~Gao, ``Ssim-motivated rate-distortion optimization for video coding,'' \emph{IEEE Transactions on Circuits and Systems for Video Technology}, vol.~22, no.~4, pp. 516--529, 2011.

\bibitem{chen2020proxiqa}
L.-H. Chen, C.~G. Bampis, Z.~Li, A.~Norkin, and A.~C. Bovik, ``Proxiqa: A proxy approach to perceptual optimization of learned image compression,'' \emph{IEEE Transactions on Image Processing}, vol.~30, pp. 360--373, 2020.

\bibitem{rouis2018perceptually}
K.~Rouis, M.-C. Larabi, and J.~B. Tahar, ``Perceptually adaptive lagrangian multiplier for hevc guided rate-distortion optimization,'' \emph{IEEE Access}, vol.~6, pp. 33\,589--33\,603, 2018.

\bibitem{mazumdar2020perceptual}
A.~Mazumdar, \emph{Perceptual Optimizations for Video Capture, Processing, and Storage Systems}.\hskip 1em plus 0.5em minus 0.4em\relax University of Washington, 2020.

\bibitem{philippe2024ed}
P.~Philippe, T.~Ladune, S.~Davenet, and T.~Leguay, ``Ed: Perceptually tuned enhanced compression model,'' \emph{arXiv preprint arXiv:2401.02145}, 2024.

\bibitem{chadha2021deep}
A.~Chadha and Y.~Andreopoulos, ``Deep perceptual preprocessing for video coding,'' in \emph{Proceedings of the IEEE/CVF Conference on Computer Vision and Pattern Recognition}, 2021, pp. 14\,852--14\,861.

\bibitem{arai2024gop}
D.~Arai, S.~Iwamura, K.~Iguchi, and A.~Ichigaya, ``Gop-based deep preprocessing for video coding,'' in \emph{2024 Picture Coding Symposium (PCS)}.\hskip 1em plus 0.5em minus 0.4em\relax IEEE, 2024, pp. 1--5.

\bibitem{yue2013cloud}
H.~Yue, X.~Sun, J.~Yang, and F.~Wu, ``Cloud-based image coding for mobile devices—toward thousands to one compression,'' \emph{IEEE Transactions on Multimedia}, vol.~15, no.~4, pp. 845--857, 2013.

\bibitem{wang2016joint}
H.~Wang, T.~Tian, M.~Ma, and J.~Wu, ``Joint compression of near-duplicate videos,'' \emph{IEEE Transactions on Multimedia}, vol.~19, no.~5, pp. 908--920, 2017.

\bibitem{mao2023peering}
Y.~Mao, P.~Chen, S.~Wang, S.~Wang, and D.~Wu, ``Peering into the sketch: Ultra-low bitrate face compression for joint human and machine perception,'' in \emph{Proceedings of the 31st ACM International Conference on Multimedia}, 2023, pp. 2564--2572.

\bibitem{oquab2021low}
M.~Oquab, P.~Stock, D.~Haziza, T.~Xu, P.~Zhang, O.~Celebi, Y.~Hasson, P.~Labatut, B.~Bose-Kolanu, T.~Peyronel \emph{et~al.}, ``Low bandwidth video-chat compression using deep generative models,'' in \emph{Proceedings of the IEEE/CVF Conference on Computer Vision and Pattern Recognition}, 2021, pp. 2388--2397.

\bibitem{tang2022generative}
A.~Tang, Y.~Huang, J.~Ling, Z.~Zhang, Y.~Zhang, R.~Xie, and L.~Song, ``Generative compression for face video: A hybrid scheme,'' in \emph{2022 IEEE International Conference on Multimedia and Expo (ICME)}.\hskip 1em plus 0.5em minus 0.4em\relax IEEE, 2022, pp. 1--6.

\bibitem{li2022towards}
B.~Li, B.~Chen, Z.~Wang, S.~Wang, and Y.~Ye, ``Towards ultra low bit-rate digital human character communication via compact 3d face descriptors,'' in \emph{2022 Data Compression Conference (DCC)}.\hskip 1em plus 0.5em minus 0.4em\relax IEEE, 2022, pp. 461--461.

\bibitem{chen2023compact}
B.~Chen, Z.~Wang, B.~Li, S.~Wang, and Y.~Ye, ``Compact temporal trajectory representation for talking face video compression,'' \emph{IEEE Transactions on Circuits and Systems for Video Technology}, vol.~33, no.~11, pp. 7009--7023, 2023.

\bibitem{chang2022conceptual}
J.~Chang, Z.~Zhao, C.~Jia, S.~Wang, L.~Yang, Q.~Mao, J.~Zhang, and S.~Ma, ``Conceptual compression via deep structure and texture synthesis,'' \emph{IEEE Transactions on Image Processing}, vol.~31, pp. 2809--2823, 2022.

\bibitem{9897729}
Z.~Wang, B.~Chen, Y.~Ye, and S.~Wang, ``Dynamic multi-reference generative prediction for face video compression,'' in \emph{IEEE International Conference on Image Processing (ICIP)}, 2022, pp. 896--900.

\bibitem{zhang2024video}
P.~Zhang, J.~Li, M.~Wang, N.~Sebe, S.~Kwong, and S.~Wang, ``When video coding meets multimodal large language models: A unified paradigm for video coding,'' \emph{arXiv preprint arXiv:2408.08093}, 2024.

\bibitem{redondi2016compress}
A.~Redondi, L.~Baroffio, L.~Bianchi, M.~Cesana, and M.~Tagliasacchi, ``Compress-then-analyze versus analyze-then-compress: What is best in visual sensor networks?'' \emph{IEEE Transactions on Mobile Computing}, vol.~15, no.~12, pp. 3000--3013, 2016.

\bibitem{weiss2008spectral}
Y.~Weiss, A.~Torralba, and R.~Fergus, ``Spectral hashing,'' \emph{Advances in Neural Information Processing Systems}, vol.~21, pp. 1753--1760, 2008.

\bibitem{chandrasekhar2009transform}
V.~Chandrasekhar, G.~Takacs, D.~Chen, S.~S. Tsai, J.~Singh, and B.~Girod, ``Transform coding of image feature descriptors,'' in \emph{Visual Communications and Image Processing 2009}, vol. 7257.\hskip 1em plus 0.5em minus 0.4em\relax International Society for Optics and Photonics, 2009, p. 725710.

\bibitem{jegou2010product}
H.~Jégou, M.~Douze, and C.~Schmid, ``Product quantization for nearest neighbor search,'' \emph{IEEE Transactions on Pattern Analysis and Machine Intelligence}, vol.~33, no.~1, pp. 117--128, 2011.

\bibitem{ge2013optimized}
T.~Ge, K.~He, Q.~Ke, and J.~Sun, ``Optimized product quantization,'' \emph{IEEE Transactions on Pattern Analysis and Machine Intelligence}, vol.~36, no.~4, pp. 744--755, 2014.

\bibitem{calonder2011brief}
M.~Calonder, V.~Lepetit, M.~Ozuysal, T.~Trzcinski, C.~Strecha, and P.~Fua, ``{BRIEF}: Computing a local binary descriptor very fast,'' \emph{IEEE Transactions on Pattern Analysis and Machine Intelligence}, vol.~34, no.~7, pp. 1281--1298, 2012.

\bibitem{rublee2011orb}
E.~Rublee, V.~Rabaud, K.~Konolige, and G.~Bradski, ``{ORB}: An efficient alternative to sift or surf,'' in \emph{2011 International Conference on Computer Vision}, 2011, pp. 2564--2571.

\bibitem{leutenegger2011brisk}
S.~Leutenegger, M.~Chli, and R.~Y. Siegwart, ``{BRISK}: Binary robust invariant scalable keypoints,'' in \emph{2011 International Conference on Computer Vision}, 2011, pp. 2548--2555.

\bibitem{zhang2014usb}
S.~Zhang, Q.~Tian, Q.~Huang, W.~Gao, and Y.~Rui, ``{USB:} ultrashort binary descriptor for fast visual matching and retrieval,'' \emph{IEEE Transactions on Image Processing}, vol.~23, no.~8, pp. 3671--3683, 2014.

\bibitem{jegou2010aggregating}
H.~Jégou, M.~Douze, C.~Schmid, and P.~Pérez, ``Aggregating local descriptors into a compact image representation,'' in \emph{2010 IEEE Computer Society Conference on Computer Vision and Pattern Recognition}, 2010, pp. 3304--3311.

\bibitem{perronnin2010large}
F.~Perronnin, Y.~Liu, J.~S{\'a}nchez, and H.~Poirier, ``Large-scale image retrieval with compressed fisher vectors,'' in \emph{2010 IEEE Computer Society Conference on Computer Vision and Pattern Recognition}.\hskip 1em plus 0.5em minus 0.4em\relax IEEE, 2010, pp. 3384--3391.

\bibitem{chen2013residual}
D.~Chen, S.~Tsai, V.~Chandrasekhar, G.~Takacs, R.~Vedantham, R.~Grzeszczuk, and B.~Girod, ``Residual enhanced visual vector as a compact signature for mobile visual search,'' \emph{Signal Processing}, vol.~93, no.~8, pp. 2316--2327, 2013.

\bibitem{chen2011residual}
D.~Chen, S.~Tsai, V.~Chandrasekhar, G.~Takacs, H.~Chen, R.~Vedantham, R.~Grzeszczuk, and B.~Girod, ``Residual enhanced visual vectors for on-device image matching,'' in \emph{2011 Conference Record of the Forty Fifth Asilomar Conference on Signals, Systems and Computers (ASILOMAR)}.\hskip 1em plus 0.5em minus 0.4em\relax IEEE, 2011, pp. 850--854.

\bibitem{lin2014rate}
J.~Lin, L.-Y. Duan, Y.~Huang, S.~Luo, T.~Huang, and W.~Gao, ``Rate-adaptive compact fisher codes for mobile visual search,'' \emph{IEEE Signal Processing Letters}, vol.~21, no.~2, pp. 195--198, 2014.

\bibitem{liu2012supervised}
W.~Liu, J.~Wang, R.~Ji, Y.-G. Jiang, and S.-F. Chang, ``Supervised hashing with kernels,'' in \emph{2012 IEEE Conference on Computer Vision and Pattern Recognition}.\hskip 1em plus 0.5em minus 0.4em\relax IEEE, 2012, pp. 2074--2081.

\bibitem{shen2015supervised}
F.~Shen, C.~Shen, W.~Liu, and H.~T. Shen, ``Supervised discrete hashing,'' in \emph{2015 IEEE Conference on Computer Vision and Pattern Recognition (CVPR)}, 2015, pp. 37--45.

\bibitem{luo2023survey}
X.~Luo, H.~Wang, D.~Wu, C.~Chen, M.~Deng, J.~Huang, and X.-S. Hua, ``A survey on deep hashing methods,'' \emph{ACM Transactions on Knowledge Discovery from Data}, vol.~17, no.~1, pp. 1--50, 2023.

\bibitem{he2024one}
L.~He, Z.~Huang, C.~Liu, R.~Li, R.~Wu, Q.~Liu, and E.~Chen, ``One-bit deep hashing: Towards resource-efficient hashing model with binary neural network,'' in \emph{Proceedings of the 32nd ACM International Conference on Multimedia}, 2024, pp. 7162--7171.

\bibitem{gu2024entropy}
L.~Gu, J.~Liu, X.~Liu, W.~Wan, and J.~Sun, ``Entropy-optimized deep weighted product quantization for image retrieval,'' \emph{IEEE Transactions on Image Processing}, vol.~33, pp. 1162--1174, 2024.

\bibitem{tu2024two}
J.~Tu, X.~Liu, Y.~Hao, R.~Hong, and M.~Wang, ``Two-step discrete hashing for cross-modal retrieval,'' \emph{IEEE Transactions on Multimedia}, vol.~26, pp. 8730--8741, 2024.

\bibitem{lin2017hnip}
J.~Lin, L.-Y. Duan, S.~Wang, Y.~Bai, Y.~Lou, V.~Chandrasekhar, T.~Huang, A.~Kot, and W.~Gao, ``{HNIP}: Compact deep invariant representations for video matching, localization, and retrieval,'' \emph{IEEE Transactions on Multimedia}, vol.~19, no.~9, pp. 1968--1983, 2017.

\bibitem{ding2017rate}
L.~Ding, Y.~Tian, H.~Fan, Y.~Wang, and T.~Huang, ``Rate-performance-loss optimization for inter-frame deep feature coding from videos,'' \emph{IEEE Transactions on Image Processing}, vol.~26, no.~12, pp. 5743--5757, 2017.

\bibitem{ding2020joint}
L.~Ding, Y.~Tian, H.~Fan, C.~Chen, and T.~Huang, ``Joint coding of local and global deep features in videos for visual search,'' \emph{IEEE Transactions on Image Processing}, vol.~29, pp. 3734--3749, 2020.

\bibitem{kang2017neurosurgeon}
Y.~Kang, J.~Hauswald, C.~Gao, A.~Rovinski, T.~Mudge, J.~Mars, and L.~Tang, ``Neurosurgeon: Collaborative intelligence between the cloud and mobile edge,'' \emph{ACM SIGARCH Computer Architecture News}, vol.~45, no.~1, pp. 615--629, 2017.

\bibitem{bragilevsky2020tensor}
L.~Bragilevsky and I.~V. Baji{\'c}, ``Tensor completion methods for collaborative intelligence,'' \emph{IEEE Access}, vol.~8, pp. 41\,162--41\,174, 2020.

\bibitem{alvar2019multi}
S.~R. Alvar and I.~V. Baji{\'c}, ``Multi-task learning with compressible features for collaborative intelligence,'' in \emph{2019 IEEE International Conference on Image Processing (ICIP)}.\hskip 1em plus 0.5em minus 0.4em\relax IEEE, 2019, pp. 1705--1709.

\bibitem{alvar2020bit}
------, ``Bit allocation for multi-task collaborative intelligence,'' in \emph{ICASSP 2020-2020 IEEE International Conference on Acoustics, Speech and Signal Processing (ICASSP)}.\hskip 1em plus 0.5em minus 0.4em\relax IEEE, 2020, pp. 4342--4346.

\bibitem{cohen2020lightweight}
R.~A. Cohen, H.~Choi, and I.~V. Baji{\'c}, ``Lightweight compression of neural network feature tensors for collaborative intelligence,'' in \emph{2020 IEEE International Conference on Multimedia and Expo (ICME)}.\hskip 1em plus 0.5em minus 0.4em\relax IEEE, 2020, pp. 1--6.

\bibitem{chen2019toward}
Z.~Chen, K.~Fan, S.~Wang, L.~Duan, W.~Lin, and A.~C. Kot, ``Toward intelligent sensing: Intermediate deep feature compression,'' \emph{IEEE Transactions on Image Processing}, vol.~29, pp. 2230--2243, 2019.

\bibitem{chen2020data}
Z.~Chen, L.-Y. Duan, S.~Wang, W.~Lin, and A.~C. Kot, ``Data representation in hybrid coding framework for feature maps compression,'' in \emph{2020 IEEE International Conference on Image Processing (ICIP)}.\hskip 1em plus 0.5em minus 0.4em\relax IEEE, 2020, pp. 3094--3098.

\bibitem{wang2021end}
S.~Wang, Z.~Wang, S.~Wang, and Y.~Ye, ``End-to-end compression towards machine vision: Network architecture design and optimization,'' \emph{IEEE Open Journal of Circuits and Systems}, vol.~2, pp. 675--685, 2021.

\bibitem{le2021image}
N.~Le, H.~Zhang, F.~Cricri, R.~Ghaznavi-Youvalari, and E.~Rahtu, ``Image coding for machines: an end-to-end learned approach,'' in \emph{IEEE International Conference on Acoustics, Speech and Signal Processing (ICASSP)}, 2021, pp. 1590--1594.

\bibitem{tian2024coding}
Y.~Tian, G.~Lu, Y.~Yan, G.~Zhai, L.~Chen, and Z.~Gao, ``A coding framework and benchmark towards low-bitrate video understanding,'' \emph{IEEE Transactions on Pattern Analysis and Machine Intelligence}, vol.~46, no.~8, pp. 5852--5872, 2024.

\bibitem{bajic2021collaborative}
I.~V. Bajić, W.~Lin, and Y.~Tian, ``Collaborative intelligence: Challenges and opportunities,'' in \emph{2021 IEEE International Conference on Acoustics, Speech and Signal Processing (ICASSP)}, 2021, pp. 8493--8497.

\bibitem{alvar2021pareto}
S.~R. Alvar and I.~V. Baji{\'c}, ``Pareto-optimal bit allocation for collaborative intelligence,'' \emph{IEEE Trans. Image Process.}, vol.~30, pp. 3348--3361, 2021.

\bibitem{chen2023transtic}
Y.-H. Chen, Y.-C. Weng, C.-H. Kao, C.~Chien, W.-C. Chiu, and W.-H. Peng, ``Transtic: Transferring transformer-based image compression from human perception to machine perception,'' in \emph{Proceedings of the IEEE/CVF International Conference on Computer Vision}, 2023, pp. 23\,297--23\,307.

\bibitem{duan2015overview}
L.-Y. Duan, V.~Chandrasekhar, J.~Chen, J.~Lin, Z.~Wang, T.~Huang, B.~Girod, and W.~Gao, ``Overview of the {MPEG-CDVS} standard,'' \emph{IEEE Transactions on Image Processing}, vol.~25, no.~1, pp. 179--194, 2015.

\bibitem{chen2014low}
J.~Chen, L.-Y. Duan, F.~Gao, J.~Cai, A.~C. Kot, and T.~Huang, ``A low complexity interest point detector,'' \emph{IEEE Signal Processing Letters}, vol.~22, no.~2, pp. 172--176, 2014.

\bibitem{paschalakis2012cdvs}
S.~Paschalakis, K.~Wnukowicz, M.~Bober, A.~Mosca, and M.~Mattelliano, ``{CDVS CE2}: Local descriptor compression proposal,'' \emph{ISO/IEC JTC1/SC29/WG11 M}, vol. 25929, 2012.

\bibitem{tsai2009location}
S.~S. Tsai, D.~Chen, G.~Takacs, V.~Chandrasekhar, J.~P. Singh, and B.~Girod, ``Location coding for mobile image retrieval,'' in \emph{Proceedings of the 5th International ICST Mobile Multimedia Communications Conference}, 2009, pp. 1--7.

\bibitem{tsai2012improved}
S.~S. Tsai, D.~Chen, G.~Takacs, V.~Chandrasekhar, M.~Makar, R.~Grzeszczuk, and B.~Girod, ``Improved coding for image feature location information,'' in \emph{Applications of Digital Image Processing XXXV}, vol. 8499.\hskip 1em plus 0.5em minus 0.4em\relax International Society for Optics and Photonics, 2012, p. 84991E.

\bibitem{duan2018compact}
L.-Y. Duan, Y.~Lou, Y.~Bai, T.~Huang, W.~Gao, V.~Chandrasekhar, J.~Lin, S.~Wang, and A.~C. Kot, ``Compact descriptors for video analysis: The emerging {MPEG} standard,'' \emph{IEEE MultiMedia}, vol.~26, no.~2, pp. 44--54, 2018.

\bibitem{yang2024video}
W.~Yang, H.~Huang, Y.~Hu, L.-Y. Duan, and J.~Liu, ``Video coding for machines: Compact visual representation compression for intelligent collaborative analytics,'' \emph{IEEE Transactions on Pattern Analysis and Machine Intelligence}, vol.~46, no.~7, pp. 5174--5191, 2024.

\bibitem{wang2019scalable}
S.~Wang, S.~Wang, X.~Zhang, S.~Wang, S.~Ma, and W.~Gao, ``Scalable facial image compression with deep feature reconstruction,'' in \emph{2019 IEEE International Conference on Image Processing (ICIP)}.\hskip 1em plus 0.5em minus 0.4em\relax IEEE, 2019, pp. 2691--2695.

\bibitem{lin2023deepsvc}
H.~Lin, B.~Chen, Z.~Zhang, J.~Lin, X.~Wang, and T.~Zhao, ``Deepsvc: Deep scalable video coding for both machine and human vision,'' in \emph{Proceedings of the 31st ACM International Conference on Multimedia}, 2023, pp. 9205--9214.

\bibitem{zhang2016real}
B.~Zhang, L.~Wang, Z.~Wang, Y.~Qiao, and H.~Wang, ``Real-time action recognition with enhanced motion vector cnns,'' in \emph{2016 IEEE Conference on Computer Vision and Pattern Recognition (CVPR)}, 2016, pp. 2718--2726.

\bibitem{ma2018joint}
S.~Ma, X.~Zhang, S.~Wang, X.~Zhang, C.~Jia, and S.~Wang, ``Joint feature and texture coding: Toward smart video representation via front-end intelligence,'' \emph{IEEE Transactions on Circuits and Systems for Video Technology}, vol.~29, no.~10, pp. 3095--3105, 2018.

\bibitem{zhang2016joint}
X.~Zhang, S.~Ma, S.~Wang, X.~Zhang, H.~Sun, and W.~Gao, ``A joint compression scheme of video feature descriptors and visual content,'' \emph{IEEE Transactions on Image Processing}, vol.~26, no.~2, pp. 633--647, 2016.

\bibitem{li2018joint}
Y.~Li, C.~Jia, S.~Wang, X.~Zhang, S.~Wang, S.~Ma, and W.~Gao, ``Joint rate-distortion optimization for simultaneous texture and deep feature compression of facial images,'' in \emph{2018 IEEE Fourth International Conference on Multimedia Big Data (BigMM)}.\hskip 1em plus 0.5em minus 0.4em\relax IEEE, 2018, pp. 1--5.

\bibitem{sun2020semantic}
S.~Sun, T.~He, and Z.~Chen, ``Semantic structured image coding framework for multiple intelligent applications,'' \emph{IEEE Transactions on Circuits and Systems for Video Technology}, vol.~31, no.~9, pp. 3631--3642, 2020.

\bibitem{wang2021towards}
S.~Wang, S.~Wang, W.~Yang, X.~Zhang, S.~Wang, S.~Ma, and W.~Gao, ``Towards analysis-friendly face representation with scalable feature and texture compression,'' \emph{IEEE Transactions on Multimedia}, vol.~24, pp. 3169--3181, 2021.

\bibitem{padamsey2022neocortex}
Z.~Padamsey, D.~Katsanevaki, N.~Dupuy, and N.~L. Rochefort, ``Neocortex saves energy by reducing coding precision during food scarcity,'' \emph{Neuron}, vol. 110, no.~2, pp. 280--296, 2022.

\bibitem{choi2022scalable}
H.~Choi and I.~V. Baji{\'c}, ``Scalable image coding for humans and machines,'' \emph{IEEE Transactions on Image Processing}, vol.~31, pp. 2739--2754, 2022.

\bibitem{10004012}
X.~Yi, H.~Wang, S.~Kwong, and C.-C. Jay~Kuo, ``Task-driven video compression for humans and machines: Framework design and optimization,'' \emph{IEEE Transactions on Multimedia}, vol.~25, pp. 8091--8102, 2023.

\bibitem{sheng2024vnvc}
X.~Sheng, L.~Li, D.~Liu, and H.~Li, ``Vnvc: A versatile neural video coding framework for efficient human-machine vision,'' \emph{IEEE Transactions on Pattern Analysis and Machine Intelligence}, vol.~46, no.~7, pp. 4579--4596, 2024.

\bibitem{ge2024task}
X.~Ge, J.~Luo, X.~Zhang, T.~Xu, G.~Lu, D.~He, J.~Geng, Y.~Wang, J.~Zhang, and H.~Qin, ``Task-aware encoder control for deep video compression,'' in \emph{Proceedings of the IEEE/CVF Conference on Computer Vision and Pattern Recognition}, 2024, pp. 26\,036--26\,045.

\bibitem{li2024human}
H.~Li and X.~Zhang, ``Human–machine collaborative image compression method based on implicit neural representations,'' \emph{IEEE Journal on Emerging and Selected Topics in Circuits and Systems}, vol.~14, no.~2, pp. 198--208, 2024.

\bibitem{zhang2024unified}
\BIBentryALTinterwordspacing
G.~Zhang, X.~Zhang, and L.~Tang, ``Unified and scalable deep image compression framework for human and machine,'' \emph{ACM Trans. Multimedia Comput. Commun. Appl.}, vol.~20, no.~10, Oct. 2024. [Online]. Available: \url{https://doi.org/10.1145/3678472}
\BIBentrySTDinterwordspacing

\bibitem{cao2023comprehensive}
\BIBentryALTinterwordspacing
Y.~Cao, S.~Li, Y.~Liu, Z.~Yan, Y.~Dai, P.~Yu, and L.~Sun, ``A survey of ai-generated content (aigc),'' \emph{ACM Comput. Surv.}, Dec. 2024, just Accepted. [Online]. Available: \url{https://doi.org/10.1145/3704262}
\BIBentrySTDinterwordspacing

\bibitem{foo2023aigc}
L.~G. Foo, H.~Rahmani, and J.~Liu, ``Aigc for various data modalities: A survey,'' \emph{arXiv preprint arXiv:2308.14177}, 2023.

\bibitem{xu2024unleashing}
M.~Xu, H.~Du, D.~Niyato, J.~Kang, Z.~Xiong, S.~Mao, Z.~Han, A.~Jamalipour, D.~I. Kim, X.~Shen, V.~C.~M. Leung, and H.~V. Poor, ``Unleashing the power of edge-cloud generative {AI} in mobile networks: A survey of {AIGC} services,'' \emph{IEEE Communications Surveys \& Tutorials}, vol.~26, no.~2, pp. 1127--1170, 2024.

\bibitem{li2024high}
B.~Li, S.~Wang, S.~Wang, and Y.~Ye, ``High efficiency image compression for large visual-language models,'' \emph{IEEE Transactions on Circuits and Systems for Video Technology}, pp. 1--1, 2024.

\bibitem{wu2019vision}
D.~Wu, Z.~Li, J.~Wang, Y.~Zheng, M.~Li, and Q.~Huang, ``Vision and challenges for knowledge centric networking,'' \emph{IEEE Wireless Communications}, vol.~26, no.~4, pp. 117--123, 2019.

\bibitem{shi2016edge}
W.~Shi, J.~Cao, Q.~Zhang, Y.~Li, and L.~Xu, ``Edge computing: Vision and challenges,'' \emph{IEEE Internet of Things Journal}, vol.~3, no.~5, pp. 637--646, 2016.

\bibitem{kuo2018data}
C.-C.~J. Kuo and Y.~Chen, ``On data-driven {Saak} transform,'' \emph{Journal of Visual Communication and Image Representation}, vol.~50, pp. 237--246, 2018.

\bibitem{luo2022semantic}
X.~Luo, H.-H. Chen, and Q.~Guo, ``Semantic communications: Overview, open issues, and future research directions,'' \emph{IEEE Wireless Communications}, vol.~29, no.~1, pp. 210--219, 2022.

\bibitem{markovic2020physics}
D.~Markovi{\'c}, A.~Mizrahi, D.~Querlioz, and J.~Grollier, ``Physics for neuromorphic computing,'' \emph{Nature Reviews Physics}, vol.~2, no.~9, pp. 499--510, 2020.

\bibitem{shastri2021photonics}
B.~J. Shastri, A.~N. Tait, T.~Ferreira~de Lima, W.~H. Pernice, H.~Bhaskaran, C.~D. Wright, and P.~R. Prucnal, ``Photonics for artificial intelligence and neuromorphic computing,'' \emph{Nature Photonics}, vol.~15, no.~2, pp. 102--114, 2021.

\end{thebibliography}

\end{document}